\documentclass[aip,preprint,amsmath,amssymb,floatfix,notitlepage,nofootinbib]{revtex4} 
\usepackage{lineno,hyperref}
 \usepackage{float}
 \usepackage[T1]{fontenc}
\usepackage{chemformula}
\usepackage{graphicx}
\usepackage{xcolor}
\usepackage{bm}
\usepackage{amsmath,amssymb}

\begin{document}

\title{Osmotically-induced rupture of viral capsids}
\author{Felipe B. M. Aguiar}
\email{felipebritoaguiar@gmail.com}
\affiliation{Instituto de F\'isica, Universidade Federal de Ouro Preto, 35400-000, Ouro Preto, MG, Brazil}
\author{Thiago Colla}
\email{colla@ufop.edu.br}
\affiliation{Instituto de F\'isica, Universidade Federal de Ouro Preto, 35400-000, Ouro Preto, MG, Brazil}

 \begin{abstract}
A simple model is proposed aimed to investigate how the amount of dissociated ions influences the mechanical stability of viral capsids.  After an osmotic and mechanical equilibrium is established with the outer solution, a non-adiabatic change in salt concentration at the external environment is considered, which results in a significant solvent inflow across the capsid surface, eventually leading to its rupture. The key assumption behind such an osmotic shock mechanism is that solvent flow takes place at timescales much shorter than the ones typical of ionic diffusion. In order to theoretically describe this effect, we herein propose a thermodynamic model based on the traditional Flory theory. The proposed approach is further combined with a continuum Hookian elastic model of surface stretching and pore-opening along the lines of a Classical Nucleation Theory (CNT), allowing us to establish the conditions under which capsid mechanical instability takes place. Despite its non-local character, the proposed model is able to capture most of the  relevant physical mechanisms controlling capsid stability, namely the volume exclusion and entropy of mixing effects among the densely-packed components, the elastic cost for capsid stretching and further pore opening, the Donnan equilibrium across the interface, as well as the large entropy loss resulting from folding the viral genome into close-packed configurations inside the capsid. It is shown that, depending on the particular combination of initial condition and capsid surface strength, the capsid can either become unstable after removal of a prescribed amount of external salt, or be fully stable against osmotic shock, regardless of the amount of  ionic dilution.\end{abstract}  

\maketitle 
\newpage 
\section{Introduction}

In spite of being one of the simplest biological structures, viruses are rather complex entities when looked upon from a physical perspective~\cite{Cas62,Tre06,Ros89,Zlo04,Roos10,Abr12,Mate13,Luque2013,Zan20,Zhd21}. This is not only due to the inherent complexity of the packed genetic material~\cite{bach14}, but also due to the large variety in structure and unique physical properties of the enveloping shell that encloses it ~\cite{Ros89,Roo07,Roos10,Luque2013,Bru15,Zhd21}. Indeed, most viruses can be simply described as a genetic material that encodes its overall functional instructions, packed onto a protecting cover, generally known as the \textit{viral capsid}~\cite{Tre06}. This capsid is a biologically designed nano-shell whose functionality encompasses the encapsulation and further transport of the viral genetic material, with the ultimate goal of bringing it intact – and further releasing it – into a host cell~\cite{Zan20,Che18,Ree19,Luque2013}. In order to accomplish such a formidable task, viral capsids are naturally endowed with a number of interesting physical properties, the most relevant of which are their semi-permeability to material flux, and their mechanical robustness against applied pressures/indentations~\cite{Ale03,Zan05}. Since the confined genetic material is comprised of a number of ionizable groups, ionic flux across the protecting shell is a key ingredient to guarantee its static equilibrium with the outer environment~\cite{Evi11,Jin12,Cor03,Col20,Zen21,Alz21}. Moreover, since most organic materials are embedded on an aqueous environment, equilibrium with different external conditions also requires water (solvent) to be freely able to flow over the capsid interface. Concomitant with such particle diffusion equilibrium, mechanical stability further demands the permeable shell to be rigid enough such as to sustain large osmotic stresses, yet keeping its structural integrity all the way to the intracell environment~\cite{Suk21}. 

Viral capsids are typically made up of a number of sub-units composed of small proteic groups, which in the early self-assembly stages combine together into small structures generally known as \textit{capsomers}. These proteic basic units (usually pentamers or hexamers) are the fundamental building blocks that self-assembly into regular closed shells ~\cite{Almendral} that encapsulate the viral genome. The dynamical process of capsid formation from the aggregation of such basic sub-units is a fascinating phenomenon that deserves special attention on its own ~\cite{Keg06,Kat10,Rij13,Mat13,Hagan14,Per15,Bru16,Twa19,Men20,Mil15,Men20,Jus20}. Capsid assembly around the packed genome is mostly driven by Coulombic forces due to their opposite charge~\cite{Keg06,Vla06,Sil08_2,Peng12}, and is in some cases further triggered by motor proteins that attach to the genome chain during the self-assembly process~\cite{Hagan14,Per15}. Capsid formation also depends on the nature of the encapsulated genome and its specific interactions with the protein groups~\cite{Gar15,Yang22}. In the case of single-strained RNA (ssRNA), capsid assembly can be further promoted by non-electrostatic interactions resulting from the branched structure of the packed genome~\cite{Tando16}. This leads to a negative pressure on the assembled structure, in strong  contrast to the case of double-strained DNA (ds-DNA) viruses, which are normally under strong stretching stress~\cite{Tando16}.  Once the capsid is formed, its stability depends on a fine equilibrium with the outer environment. The understanding and control of the various physical mechanisms that dictate such stability under different external conditions is a rather challenging task~\cite{Zhd21}. This is not only due to the aforementioned peculiar properties assigned to the confining shell~\cite{Ros89,Cas13}, but also due its sensibility to external conditions.

In a typical coarse-graining framework, the viral genome can be represented as a long linear chain molecule comprising a very large number of elementary entities (either nucleotide or base-pairs for single or double strained genome, respectively) which are coarse-grained into connected beads of a polymer chain~\cite{Mar17}. Typically, the genome length is way larger than the capsid dimensions, forcing the encapsulated chain to be arranged into strongly compact configurations~\cite{Spa05}. Keeping such coil-like conformations of the confined genome requires a large elastic cost~\cite{Hir15,Vet15,Rap16,Mar17}, and also results in huge entropy loss~\cite{flory}. Furthermore, the elementary RNA units contain functional phosphate groups, which become ionizable in the underlying aqueous environment, thus acquiring a net charge and releasing their own counterions into the outer environment~\cite{Lev02,Lev96,Kuh98,Sil12}. Therefore, the close-packed configuration of such a charged linear chain also entails an extra energetic penalty due to strong electrolyte repulsion between neighboring charged groups~\cite{Tzli03,Pet07,Jeh15,Cor03}. In such a densely-packed inner environment, exclusion volume effects among the different confined particles also leads to a large entropy loss, for the confined particles have no room to explore a large variety of internal configurations~\cite{Smi14}. All these contributions manifest themselves as a strong pressure at the capsid inner walls, in an attempt to swell the whole structure.  In some classes of viruses, bacteriophages in particular, such large internal pressures play a key role in driven genome ejection into the host cell~\cite{Mee08,Bran19,Ian01,Roo07}.

There are several situations in which changes in the capsid surroundings lead to its mechanical instability~\cite{Evi08,Jon11,Evi11,Chen12,Liu21,Qiu12}. Given the importance in controlling viral stability, many works have been recently devoted to elucidating the physical mechanisms dictating capsid formation and its mechanical stability in a given environment~\cite{Keg06,Roo07,Evi11,Bru16,Che18}. Unfortunately, the huge variety of sizes, shapes and inner structures inherent to these objects requires the use of complex theoretical tools, yet aimed at the description of particular types of viruses subject to specific external conditions. On the other hand, simple approaches designed to describe a wider class of systems usually lack accuracy when applied to case-specific situations. However, such generic models are useful in providing qualitative estimations and, above all, valuable insights on the relevant  mechanisms controlling capsid stability, as well as on the interplay between different parameters taking part on this process~\cite{Cer22}. Recently, one such an approach has been proposed to investigate the mechanical properties of empty capsids in the framework of an elastic shell model, along the lines of a classical nucleation theory~\cite{Col20}. Our main goal in the present work is to apply a similar approach to study the stability of capsids loaded with a large, close-packed cargo, representing the ssRNA genome of a viral capsid. The proposed model puts together various approaches that incorporate some of the key mechanisms behind capsid stability: the Donnan model for charge flow across the interface, a Flory approach to describe both entropic and size effects for the mixture of chain-like and free monomers~\cite{flory,gennes,pol_phys,int_pol_phys} together with a Hookian isotropic model for shell deformation. These tools are then combined with a classical nucleation theory (CNT) that describes capsid stability against disintegration~\cite{Lev04,Idi04,Levin04}. Particular emphasis is placed on the so-called osmotic shock mechanism~\cite{And50,Lei66,Cor03}, in which capsid rupture is driven by a sudden (\textit{i. e.}, non-adiabatic) dilution of external ionic concentration, assuming that solvent has much faster diffusion times in comparison to ionic ones. As we shall see, this might result in a  very large solvent intake by the capsid, eventually leading to its irreversible rupture.

The remaining of this paper is organized as follows. In Section II, the model system under investigation and some of its general phenomenological aspects are outlined. Next, in Section III, the proposed theoretical approach is described, along with some brief discussion of its key predictions. Section IV then is devoted to present some general results for a selected class of system parameters. Finally, concluding remarks and perspective for further investigations are discussed in Section V. Some of the more specific analysis underlying the proposed theoretical approach are then left for the appendices.

\section{Model system}

\begin{figure}[h!]
\centering
\includegraphics[width=7.5cm,height=7cm]{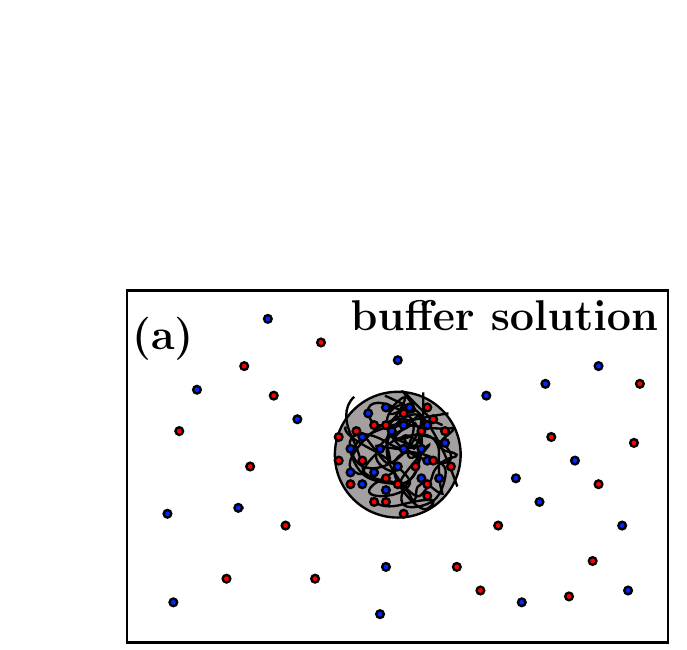}\hspace{0.3cm}
\includegraphics[width=7cm,height=3.75cm]{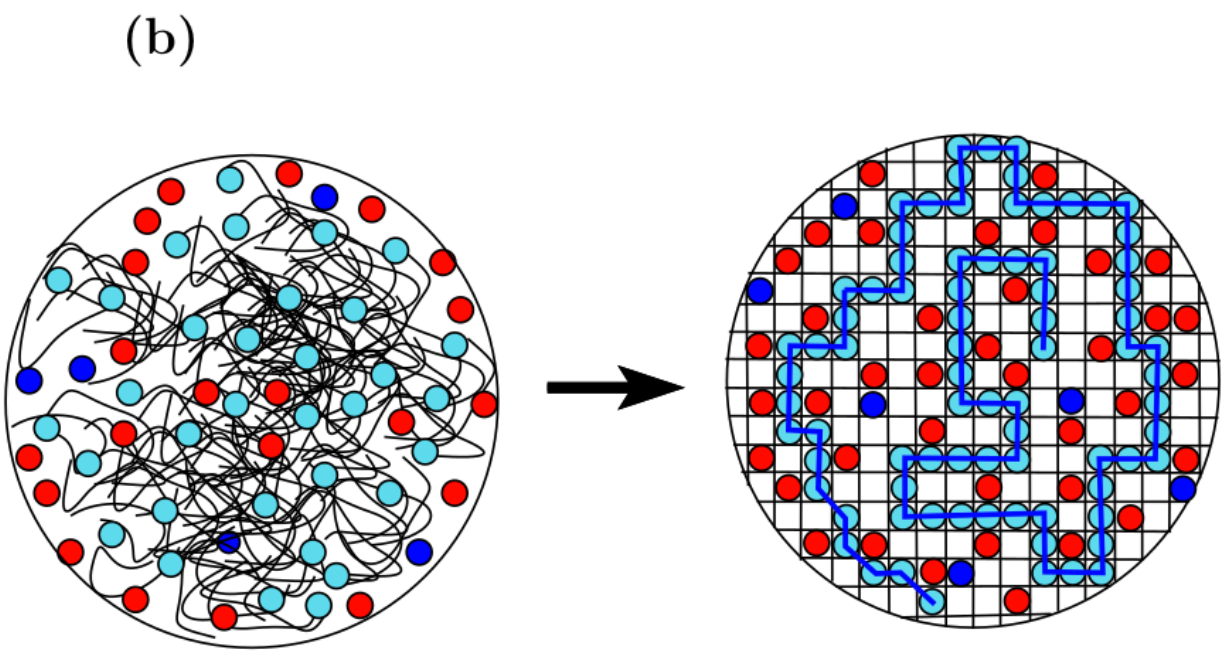}
\caption{Sketch of our model system. In (a), a spherical capsid of radius $R$ is in both osmotic and mechanical equilibrium with its outer surroundings, which are represented as a buffer solution with averaged ionic concentration $c_s$. In (b), the employed lattice model geometry is depicted. The spherical volume of the capsid is partitioned into a regular cubic lattice. The sizes of the lattice sites are chosen such as each site is able to accommodate one single particle, which can be a cation (red spheres), an anion (dark blue spheres) or a connected chain monomer (light-blue spheres). }
\label{fig:fig1}
\end{figure}

The family of viruses we aim to describe here are the ones comprised of ssRNA confined into charged capsids bearing quasi-spherical symmetry. It is important to note that the main properties of these viruses are quite distinct from those made up dsDNA, specially regarding their genome length, flexibility, self-assembly and loading/ejection mechanisms~\cite{Tando16}. To model these systems, we consider a spherical, structureless and infinitely thin viral capsid of unstretched radius $R_0$, embedded on an aqueous environment (representing the external buffer solution) containing a given concentration of dissociated, monovalent electrolyte. The capsid cargo comprises the viral genome, along with free ions dissolved and solvent molecules (see Fig.~\ref{fig:fig1}a). 

\begin{figure}[h!]
\centering
\includegraphics[width=9cm,height=6cm]{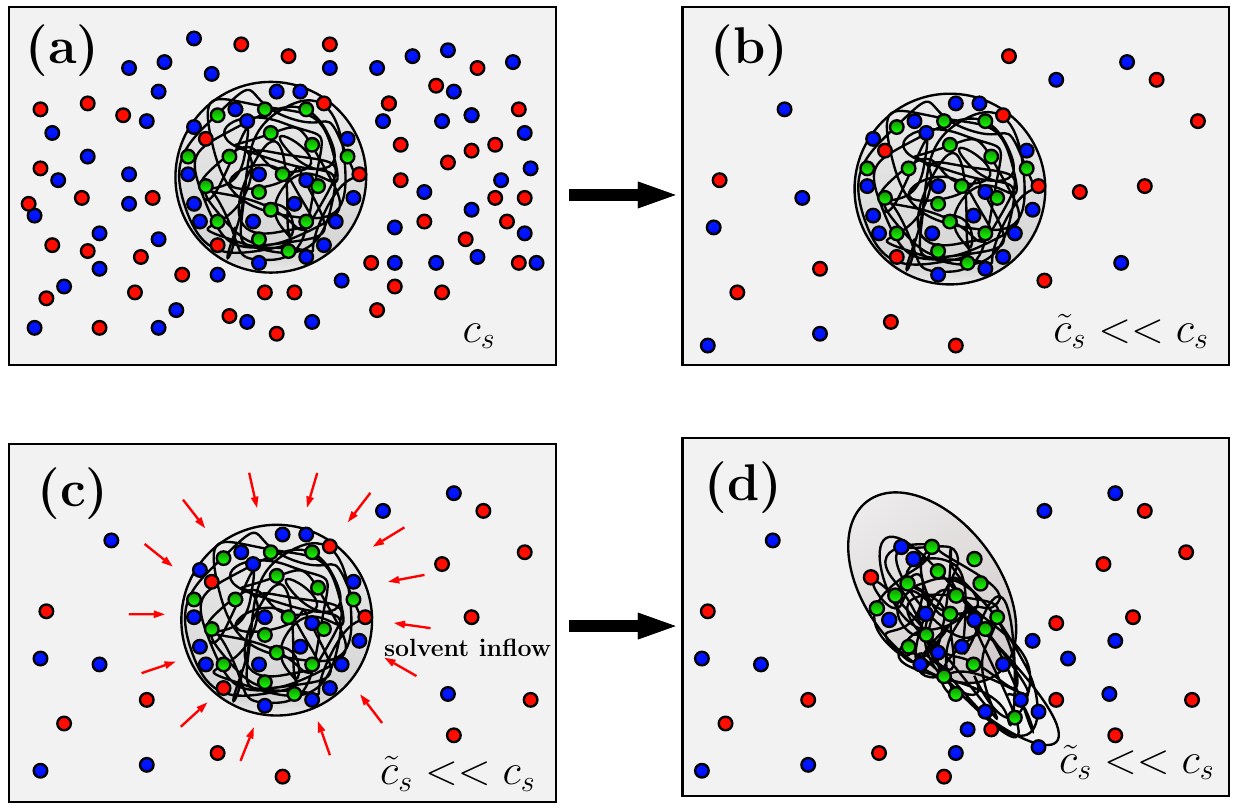}
\caption{Main steps towards an osmotic shock mechanism of a viral capsid. (a) The stable capsid is in osmotic equilibrium with a buffer solution of salt concentration $c_s$. (b) The surrounding ions undergo a quick, non-adiabatic dilution to a new concentration $\tilde{c}_s\ll c_{s}$. (c) In such a short time scale, a considerable ionic diffusion is not able to take place. As a consequence, a large solvent inflow into the capsid will be established in order to restore osmotic equilibrium with the external environment. (d) If the resulting solvent flux is large enough, the overloaded capsid will no longer be able to withstand the strong stretching force on its inner surface. The mechanical structure then collapses, leading to the leakage of its internal material out to the buffer solution.  }
\label{fig:fig2}
\end{figure}

At first, we assume that both ionic species and solvent are free to move across the shell, until an equilibrium situation of vanishing net flux is achieved. As we are dealing with ssRNA, the dominant contribution from chain confinement comes from the entropy loss resulting from folding the long linear chain into close-packed configurations, which is in this case much larger than the associated elastic energy cost~\cite{gennes,Fred05}. The polynucleotide is coarse-grained into a sequence of freely-jointed segments representing individual nucleotides, whose size thus corresponds to a typical distance between neighboring nucleotides along the chain. We consider a total of $M$ such connected beads over the sequence, each bearing a radius of $r_m=0.3$~nm -- a value within the typical size range ($\sim 0.2$ -- $0.37$~nm) employed in coarse-graining molecular approaches of ssRNA chains ~\cite{Zha04,Men11,Kim15,Zhe12,Chi13}. Usual values of chain length $M$ characteristics of polynucleotides representing single-strained genome lie mostly within the range from $\sim 10^3$ to $\sim 2\times 10^4$ nucleotides~\cite{Cam15,Chai19}. We will henceforth restrict our analysis to this specific range of values.  The system as a whole (capsid plus buffer solution) contains a total of $N_{+}$ cations and $N_{-}$ anions of radii $r_{ion}=0.2$~nm (typical of hydrated ions in aqueous solutions~\cite{Lev02}). In equilibrium, $N^{0}_{\pm}$ of such ions will be confined inside the capsid. The buffer solution contains dissolved ions of concentration $c_s=(4\pi r_{ion}^3\phi_s)/3$, where $\phi_s$ is the corresponding packing fraction. In addition, the system also contains dissociated counterions (anions) from the capsid, which bears a surface charge of $\sigma_c=0.4e/\text{nm}^2$, where $e$ is the elementary charge. This value corresponds to the upper bound of the estimated surface charge of typical viral capsids~\cite{Los12}. The equilibrium volume of the capsid, $V_0$, is normally overcrowded by the $M$ connected beads, alongside with $N^{0}_++N^{0}_{-}$ confined ions; the remaining $V_0-M-N^{0}_--N^{0}_{+}$ free internal size is considered to be occupied by solvent molecules. In order to account for strong exclusion volume effects in such a highly compact environment, we herein employ a Flory theory of linear chains~\cite{flory}. The confined space inside the capsid is divided into a very large number of lattice sites, whose size scales with a typical particle size (see Fig. \ref{fig:fig1}b). Each lattice site is constrained to be occupied by one single particle at most (either dissolved ion or polymer bead). In this framework, empty lattice sites represent the background solvent. The numbers of condensed ions $N^{0}_{\pm}$ in equilibrium are not independent, since the confined fluid is constrained to fulfill the electroneutrality condition,
\begin{equation}
z_{+}N^0_{+}-z_{-}N^0_{-}-z_mM+Z_c=0,
\label{neut}
\end{equation}
where $Z_c=4\pi R_0^2\sigma_c$  is the number of (positively charged) ionized charged groups attributed to the capsid surface, $z_{-}=z_{-}=z_{m}=1$ are anion, cation and monomer charges. This condition leads to the building-up of a potential difference across the interface, widely known in the physical-chemistry community as the Donnan potential~\cite{Tam98,Odi03,Tin11}. The equilibrium concentrations will be the ones that minimize the overall free-energy under the constrain (\ref{neut}). This osmotic equilibrium condition will be further combined with a mechanical equilibrium to obtain the equilibrium size of the stretched capsid. Then, a second scenario will be considered to describe the osmotic shock mechanism: the ionic concentration at the buffer suddenly drops considerably, within a time interval much shorter than the typical ionic diffusion timescale. It is then assumed that solvent flow across the capsid takes place at much shorter time intervals. As a result, a transient state is established shortly after ionic dilution, in which only solvent molecules are able to diffuse across the capsid in order to recover osmotic equilibrium. This leads to a significantly swelling of the capsid, eventually resulting resulting in capsid rupture and material ejection, as sketched in Fig.~\ref{fig:fig2}.

\section{Thermodynamic model of a loaded capsid}

The model system designed above will now be used to investigate osmotic and mechanical stability of capsids under different external conditions, along with the analysis of capsid stability against osmotic shock.  

\subsection{Osmotic equilibrium}

We start by considering a situation in which the capsid contains $N_{+}$ and $N_{-}$ condensed cations and anions. Let ${N}'_{+}$ and ${N}'_{-}$ be the corresponding number of dissolved ions at the buffer solution. Overall particle conservation requires that $N_{+}+N'_{+}={N}_{s}$ and $N_{-}+N'_{-}=N_s$, where $N_s$ denotes the total number of dissolved ions across the entire system. Likewise, the total system volume $\mathcal{V}$ is the sum of the capsid volume, $V$, and the volume of the outer buffer $V'$, such that $\mathcal{V}=V+V'$ is fixed. Neither the number of condensed ions nor the capsid volume are known \textit{a priori}, and their equilibrium values should be the ones that minimize the total free energy under the constraint of particle and volume conservation. The free energy can be split into external and internal contributions,
\begin{equation}
\beta\mathcal{F}=\beta\mathcal{F}_{in}+\beta\mathcal{F}_{ex},
\label{f_tot1}
\end{equation}
 where $\beta=1/k_BT$ is the inverse thermal energy, $k_B$ is the Boltzmann constant and $T$ the bath temperature. Local charge neutrality is assumed all over the system. As a consequence, the free energy at the outer buffer, $\beta\mathcal{F}_{ex}$, comprises mixing entropy contributions alone, which can be readily computed as (see Appendix A):
\begin{equation}
\beta\mathcal{F}_{ex}=N'_{+}\ln(\phi_+')+N_{-}'\ln(\phi_{-}')+(V'-N_{+}'-N_{-}')\ln(1-\phi_{+}'-\phi_{-}').
\label{f_ex1}
\end{equation}
The first two terms on the r.h.s of this relation represent ideal free energies of mixing of cations and anions, respectively, whereas the last term stands form the solvent entropic contributions. The quantity $V'$ above is to be identified with total number of available lattice sites at the outer solution. Here, $\phi'_{i}=\dfrac{N_i'}{V'}$ denotes the overall external packing fraction of the $i$-th component. Note that the primed quantities are implicit functions of the thermodynamic variables of the confined system through the relations $V'=\mathcal{V}-V$ and $N'_{\pm}=N_s-N_{\pm}$. Apart from these mixing ideal contributions, the free energy of the inner system also comprises an entropic, self-avoiding contribution from the confined polyelectrolyte, which in the context of a Flory-like approach can be written as (see Appendix A)
\begin{equation}
\beta \mathcal{F}_{chain}=(V-M)\ln(1-\phi_m)-M[\ln(\ell-1)-1],
\label{fc}
\end{equation}
where $\phi_m=\dfrac{M}{V}$ is the averaged packing fraction of the encapsulated genome, and $\ell$ represents the lattice coordination number. Since the number of confined monomers $M$ is throughout constant, the last term adds a free energy constant with no physical relevance in the present context. Note that the capsid volume $V$ is to be identified, in the lattice model, as the total number of available sites within the capsid matrix. We assume that the chain is free to explore to whole interior of the capsid. In practice, electrostatic interactions with the charged surface should futher confine the ssRNA chain within an annular region close to the interface, amounting to an extra entropy loss~\cite{Scho13}. On the other hand, annealed branching of the ssRNA segments (absent in the linear-chain description) is known to favor compact structures of the confined chains, thus lowering the chain free-energy~\cite{Schwab09,Roya09,Scho13}. Scaling arguments can be in principle employed to incorporate these effects into the Flory theory~\cite{Roya09}.

In addition to the above contribution from the confined chain, the internal free energy also contains contributions from the ideal mixing free energy of absorbed ions, which are given by (see Appendix A)
\begin{equation}
\begin{split}
\beta\mathcal{F}_{ion}=(V-M-N_{+}-N_{-})\ln(1-\phi_m-\phi_{-}-\phi_{+})-N_{+}\ln(\phi_{+})-N_{-}\ln(\phi_{-}).\\
-(V-M)\ln(1-\phi_m)
\end{split}
\label{f_ions}
\end{equation}

Combining this contribution with the ideal free energy of the confined chain (\ref{fc}) yields
\begin{equation}
\beta\mathcal{F}_{in}=N_{+}\ln(\phi_+)+N_{-}\ln(\phi_-)-M[\ln(\ell-1)-1]+(V-M-N_+-N_-)\ln(1-\phi_+-\phi_--\phi_m).
\label{f_in}
\end{equation}  
The last term above represents the mixing free energy associated to the solvent mobility. The osmotic equilibrium condition can be obtained applying the Euler-Lagrange condition $\dfrac{\partial \mathcal{F}}{\partial N_{\pm}}\biggr\arrowvert_{N^0_{\pm}}=0$, together with (\ref{f_tot1}), (\ref{fc}) and (\ref{f_in}). However, as pointed out earlier, the particle numbers $N_{\pm}$ of confined ions are not mutually independent, since they are bound to satisfy the electroneutrality condition, (\ref{neut}). To account for this requirement, we introduce a Lagrange multiplier $\mu_D$ that enforces condition (\ref{neut}) while applying the stationary condition. Therefore, the free energy to be minimized reads as
\begin{equation}  
\begin{split}
\beta\mathcal{F}& =N'_{+}\ln(\phi_+')+N_{-}'\ln(\phi_{-}')+(V'-N_{+}'-N_{-}')\ln(1-\phi_{+}'-\phi_{-}')+\mu_D(z_+N_+-z_{-}N_{-}-z_mM+Z_{c})\\
&+N_{+}\ln(\phi_+)
+N_{-}\ln(\phi_-)+(V-M-N_+-N_-)\ln(1-\phi_+-\phi_--\phi_m)-M[\ln(\ell-1)-1].
\end{split}
\label{f_tot2}
\end{equation} 
The Lagrange multiplier $\mu_D$ can be identified with the Donnan potential. Application of the stationary condition can now be readily employed, considering $N_{\pm}$ as independent variables, and noticing that $\dfrac{\partial}{\partial N_{\pm}}=-\dfrac{\partial}{\partial N'_{\pm}}$ in virtue of the particle conservation constraint. The Euler-Lagrange condition thus leads to the following equilibrium distributions of confined ions:
\begin{equation}
\phi^0_{\pm}=\dfrac{(1-\phi_m)\phi_s}{1+\phi_{s}(e^{-\beta z_{+}\mu_D}+e^{\beta z_{-}\mu_D}-2)}e^{\mp \beta z_{\pm}\mu_D},
\label{phi_eq1}
 \end{equation}
where we have used $N_s\gg N_{\pm}$ and  $\mathcal{V}\gg V$, such that  $N'_{\pm}/V'=\dfrac{(N_s-N_{\pm})}{(\mathcal{V}-V)}=\dfrac{N_s}{\mathcal{V}}\left(\dfrac{1-N_{\pm}/N_s}{1+V/\mathcal{V}}\right)\approx \phi_s$. This means that the buffer solution acts in practice as a particle reservoir to the confined system.  Indeed, the Euler-Lagrange condition for the equilibrium concentrations amounts to the requirement that chemical potentials of the confined electrolyte must equal the ones at the buffer solution. Note that the equilibrium packing fractions in (\ref{phi_eq1}) written in terms of the (yet undetermined) Lagrange multiplier $\mu_D$. This quantity can be computed by inserting (\ref{phi_eq1}) into the charge-neutrality condition (\ref{neut}). For the case of a symmetric electrolyte, $z_{\pm}=\pm z$, the following relation is obtained:
\begin{equation}
\cosh(\beta\mu_D) = \dfrac{\eta\sqrt{\left(1+\dfrac{Z_c\eta}{zV(1-\phi_m)}\right)^2(1-4\phi_s)+(2\phi_s\eta)^2}+\left(1+\dfrac{Z_c\eta}{zV(1-\phi_m)}\right)^2(1-2\phi_s)}{2\phi_s\left[\eta^2-\left(1+\dfrac{Z_c\eta}{zV(1-\phi_m)}\right)^2\right]},
\label{mu_D1}
 \end{equation}
where the parameter $\eta$ is defined as $\eta\equiv \dfrac{z(1-\phi_m)}{V(z_m\phi_m-Z_cv_{ion})}$, and $v_{ion}=4\pi r_{ion}^3/3$ is the ionic volume. In virtue of the combined finite size and charge neutrality effects, the parameter $\eta$ must always fulfill $|\eta|\ge 1$. The algebraic relation above can be explicitly solved for $\mu_D$, resulting in the following expression for the Donnan potential:
\begin{equation}
\mu_D=\ln\left(\dfrac{1-2\phi_s}{2\phi_s(\eta+1)}\right)+\ln\left[\sqrt{1+\left(\dfrac{2\phi_s}{1-2\phi_s}\right)^2\left(\eta^2-1\right)}-1\right].
\label{mu_D2}
\end{equation}
The Donnan potential features a slow, logarithm divergence, $\mu_D\approx \ln(\phi_s)$, as $\phi_s\rightarrow 0$. In this case, a very strong electric field must be established across the interface in order to sustain the sharp drop in ionic concentrations resulting from the electroneutrality requirement. The electrostatic contributions in such low ionic strengths thus play a dominant role in determining the mechanical equilibrium of a stressed capsid. Substitution of the Donnan potential from (\ref{mu_D2}) in (\ref{phi_eq1}) leads to the following expression for the equilibrium distribution of confined ions in terms of the capsid volume $V$:
\begin{equation}
\phi^0_{\pm}=\dfrac{(1-\phi_m)(1-2\phi_s)}{2(\eta+1)}\dfrac{\left[\eta^2-\left(1+\dfrac{Z_c\eta}{zV(1-\phi_m)}\right)^2\right]\left[\left(\sqrt{1+\left(\dfrac{2\phi_s}{1+2\phi_s}\right)^2(\eta^2-1)}-1\right)\right]^{\mp 1}}{\eta^2(1-2\phi_s)+\eta\sqrt{\left(1+\dfrac{Z_c\eta}{zV(1-\phi_m)}\right)^2(1-4\phi_s)+(2\phi_s\eta)^2}}.
\label{phi_eq2}
\end{equation}

\subsection{Mechanical stability}

So far the capsid's size has been held constant (recall that minimization with respect to particle numbers was carried out at constant volume). Since the chain length $M$ is also fixed, the fraction of condensed solvent can be computed as $\phi_w=(V-N^0_{+}-N^0_{-}-M)/V=1-\phi_+-\phi_--\phi_m$. Using the above relations for the equilibrium distributions, one finds
\begin{equation}
\phi_w=\dfrac{(1-\phi_m)\left[1-\left(\dfrac{1}{\eta}+\dfrac{Z_c}{zV(1-\phi_m)}\right)^2\right]}{1+\sqrt{\left(\dfrac{1}{\eta}+\dfrac{Z_c}{zV(1-\phi_m)}\right)^2+\left(\dfrac{2\phi_s}{\eta(1-2\phi_s)}\right)^2}}.
\label{phi_w}
\end{equation}
Once osmotic equilibrium is settled, a difference in pressure is developed across the inner and outer sides  of the capsid (\textit{i. e.}, a non-vanishing osmotic pressure is established). Since the capsid surface is flexible, size fluctuations are naturally allowed to take place, and the volume of the confined system can undergo small variations. The equilibrium condition that dictates such size fluctuations is the one in which surface deformations induced by the osmotic pressure should be exactly counter-balanced by the elastic response of the stressed surface. In the context of the proposed lattice model, size fluctuations can be directly assigned to an inward (outward) solvent flow, which leads to the swelling (de-swelling) of the capsid inner volume. In order to account for this effect, we should consider the capsid size as an addition variational parameter, to be optimized under the constraint of fixed volume of the entire compartment, $\mathcal{V}=V+V'$, with $V'\gg V$. To this end, an elastic penalty due to the concomitant stretching/compression of the surface should be considered together with the free energy contributions in (\ref{f_tot1}). Since viral capsids are in general rigid objects, the small deformations in the uniform surface can be accounted using a simple, Hooke-like elastic contribution:
\begin{equation}
\beta\mathcal{F}_{sur}=\beta\kappa\dfrac{(A-A_0)^2}{2A_0},
\label{F_sur}
\end{equation}
where $\kappa$ is the elastic modulus of the capsid, $A$ denotes its surface area after deformation, $A_0$ being the unstressed surface area. This elastic energy cost presumes that the surface strains are both isotropic and small. Despite the inherent inhomogeneity in shape of viral capsids (which usually precludes the case of uniform deformations), good estimations for the order of magnitude of the elastic parameter $\kappa$ can be obtained from indentation experiments~\cite{Roo07} and theoretical approaches~\cite{Los13}. Combining (\ref{f_tot1}) and (\ref{F_sur}) the stationary condition for the total free energy in terms of volume variations is 
\begin{equation}
\dfrac{\partial }{\partial V}\left(\mathcal{F}_{in}+\mathcal F_{ex}+\mathcal{F}_{sur}\right)=0\Longrightarrow \dfrac{\partial\mathcal{F}_{in}}{\partial V}-\dfrac{\partial\mathcal{F}_{ex}}{\partial{V}'}=-\dfrac{\partial\mathcal{F}_{sur}}{\partial V},
\label{osm1}
\end{equation}
where we used the fixed volume condition $\mathcal{V}=V+V'$. The derivatives above are nothing but the inner ($P_{in}=-\frac{\partial \mathcal{F}_{in}}{\partial V}$) and outer ($P_{ex}=-\frac{\partial \mathcal{F}_{ex}}{\partial V'}$) pressures acting on the capsid interface. Modeling the capsid as a spherical object of radius $R$ ($A=4\pi R^2$) the condition above can be expressed as
\begin{equation}
\beta\Pi = -\dfrac{2\beta\kappa}{R}\left(\dfrac{A}{A_0}-1\right),
\label{pi_el}
\end{equation}
where $\Pi =  P_{in}-P_{ex}$ defines the osmotic pressure across the capsid surface. The osmotic pressure can be computed \textit{via} the volume derivative of (\ref{f_ex1}) and (\ref{f_in}) taken at constant number of ions inside ($N^0_{\pm}$) and outside the capsid ($N'_{\pm}=N_s-N^0_{\pm}$), fixed at their equilibrium values. Such volume changes at constant number of ions can be interpreted as a solvent flow into/out the capsid. The external pressure reads as
\begin{equation}
\beta P_{ex}=-\ln\left(1-\phi'_{+}-\phi'_{-}\right)=-\ln(1-2\phi_s).
\label{P_ex}
\end{equation}
In the last equality above, we used the fact that $\phi'_{\pm}=\phi_s$ in the limit when $V'\gg V$. Since maximum close-packing condition demands $2\phi_s<1$, it follows that the external pressure is always positive, and therefore its contribution to the osmotic pressure $\beta\Pi=\beta P_{in}-\beta P_{ex}$ is always negative, \textit{i. e.}, it provides an inward force which in all cases attempts to shrink the surface. Besides, it grows indefinitely in magnitude as the limit where the highest packing $\phi_s\rightarrow 1/2$ is approached. Finally, we notice that, in the limit of small ionic buffer concentrations ($\phi_s\ll 1$), exclusion volume effects can be neglected, and the limit of an ideal gas of point-like particles, $\beta P=2\phi_s$ , is recovered~\footnote{In such limit of vanishing particle sizes, the packing fractions can be directly identified with the overall particle concentrations.}. Similarly, the internal pressure can be computed from the volume derivative of (\ref{f_in}):
\begin{equation}
\beta P_{in}=-\ln\left(1-\phi_m-\phi^0_{+}-\phi^0_{-}\right)-\phi_m=\ln(\phi_w)-\phi_m.
\label{P_in1}
\end{equation} 
 The internal contribution to the osmotic pressure displays a unbounded growth at solvent concentrations ($\phi_w\rightarrow 0$). When such a  close-packing condition is fulfilled, a very strong osmotic pressure is established across the capsid, resulting in a large solvent inward flux, and capsid swelling. In the limit of very small concentrations -- when size effects become negligible -- the limit $\ln(1-x)\approx x$ leads to the well-known result $\beta P_{in}\approx  \phi^0_++\phi^0_-$ for the internal contribution of a system of point-like ions in the ideal-gas limit. Combining (\ref{phi_w}), (\ref{P_ex}) and (\ref{P_in1}), the mechanical equilibrium condition (\ref{osm1}) can be further expressed in terms of fixed macroscopic quantities:
\begin{equation}
\beta \Pi = \ln\left[\dfrac{(1-\phi_m)\left(1-\left(\dfrac{1}{\eta}+\dfrac{Z_c}{zV(1-\phi_m)}\right)^2\right)}{1+\sqrt{\left(\dfrac{1}{\eta}+\dfrac{Z_c}{zV(1-\phi_m)}\right)^2+\left(\dfrac{2\phi_s}{\eta(1-2\phi_s)}\right)^2}}\right]-\phi_m=-\dfrac{2\kappa}{R}\left(1-\dfrac{A_0}{A}\right),
\label{pi_1}
\end{equation} 
so equilibrium volume $V$ can be obtained and all relevant quantities dictating the equilibrium of the capsid with its external environment are known in terms of the genome size $M$, the capsid charge $Z_c$ and elastic modulus $\kappa$, as well as the external salt concentration $\phi_s$. It only remains to determine whether this equilibrium condition will be stable against mechanical rupture of the stretched capsid. To address this issue, we now consider the possibility of a pore opening (\textit{i. e.}, a local capsomer disassembly). Pore opening will partially relieve the capsid elastic energy resulting from surface stretching. On the other hand, opening a surface hole requires the breakage of molecular bounds holding the capsomers together, thus demanding an extra energy cost. Mechanical stability will be dictated by a fine balance between these competing contributions. These effects are accounted for by considering the energy change $\Delta U$ resulting from pore opening. In a mean-field approach, the energy cost for the opening of a surface disc of radius $r$ (with $r\ll R$) can be expressed as $U_p\approx 2\pi r\gamma$, where $\gamma$ defines a line tension of the capsid, proportional to the bounding energy keeping the protein groups together. The energy required for opening a pore of radius $r$ thus reads as
\begin{equation}
\Delta U(r)=\dfrac{\kappa}{2A_0}[(A-A_0)^2-(\Delta A-A_0)^2]+2\pi r\gamma,
\label{del_U}
\end{equation}
 where $\Delta A = A-\pi r^2$ is the surface area after a hole of size $r$ is created. The first term on the r.h.s represents the decrease in elastic energy upon pore opening, while the second term is the energy cost for bound breakage all along the exposed pore rim. In cases of not too small deformations, $R/R_0> 1$ (where $R$ is the radius of the stretched capsid), the function $\Delta U(r)$ is a monotonically increasing function of $r$, indicating that pore formation is always unfavorable. However, at very small deformations ($R\gtrsim R_0$), the energy difference $\Delta U(r)$ needed for pore opening displays a non-monotonic behavior: it undergoes a local maximum at small pore sizes, before growing up indefinitely at larger $r$ (see Appendix B). Such energy barrier is a usual feature in classical nucleation theories~\cite{Blaa04}. If a thermal fluctuation is large enough to overcome the energy barrier corresponding to this local maximum (the so-called nucleation barrier), a new equilibrium state related to the next local minimum will be achieved, corresponding to stable opening of a surface pore. What happens next depends crucially on the dynamical stability of the pore, as particle diffusion might either favor pore closure or leads to its further opening -- when irreversible rupture takes place. In this work, we will not be concerned about the dynamical stability of the opened hole. We will rather assume that, once pore opening becomes favorable, the capsid will loose its mechanical stability against pore formation and further rupture. Following the proposed nucleation scenario, the stable region will be the one in which $\Delta U(r)$ displays no extremes for $r>0$ (\textit{i. e.}, it increases monotonically at all pore sizes). This will happen whenever the size of the stretched capsid does not exceed a limiting threshold value $R=R_{crit}$, given by (see Appendix B)
\begin{equation}
R_{crit} =  R_0\sqrt{1+\dfrac{3}{4}\left(\dfrac{4\gamma^2}{\kappa^2R_0^2}\right)^{1/3}}.
\label{R_crit}
\end{equation}
This quantity depends only on the underlying elastic constants, as well as on the unstressed capsid radius $R_0$. It can thus be computed regardless of the equilibrium particle numbers and equilibrium sizes, which are the natural outputs of the proposed thermodynamic lattice model. Such a convenient decoupling between thermodynamic equilibrium and mechanical stability allows one to easily identify the combinations of thermodynamic parameters which lead to equilibrium capsid sizes smaller than the limiting value $R_{crit}$ shown above -- for which mechanical stability can be guaranteed.  

\subsection{Osmotic shock}

Having established a theoretical framework for describing osmotic and mechanical equilibrium under various external conditions, we are now able to address the question of capsid stability against osmotic shock. To this end, we simply have to consider a new mechanical equilibrium condition, a situation where the external ionic strength undergoes an abrupt reduction. We assume that the osmotic dynamics is such that no considerable ionic flow into capsid  can be established within such small time scales. Solvent molecules, on the other hand, do have time to diffuse across the capsid. This corresponds to allowing for volume changing while keeping the number of condensed ions constant,  precisely the assumption behind mechanical equilibrium that leads to (\ref{pi_1}). We suppose that the external salt concentration changes from $\phi_s$ to a new packing fraction $\tilde{\phi}_s$ (with $\tilde{\phi}_s<\phi_s$), retaining the number of ions inside the capsid. This implies that the internal pressure is still described by (\ref{P_in1}), while the external pressure retains its shape given by (\ref{P_ex}) -- now with $\phi_s$ replaced by the new external ionic packing fraction $\tilde{\phi}_s$. Accordingly, the internal ionic fractions are $\tilde{\phi}^{0}_{\pm}=N^{0}_{\pm}/\tilde{V}$, where $\tilde{V}$ is the new capsid size, and $N^{0}_{\pm}$ are the number of condensed ions before osmotic shock is enforced. Under such replacements, the mechanical equilibrium condition in (\ref{pi_1}) now goes to
\begin{equation}
\beta \Pi_{shock} = \ln\left[\dfrac{(1-\tilde{\phi}_m)\left(1-\left(\dfrac{1}{\tilde{\eta}}+\dfrac{Z_c}{z\tilde{V}(1-\tilde{\phi}_m)}\right)^2\right)}{1+\sqrt{\left(\dfrac{1}{\tilde{\eta}}+\dfrac{Z_c}{z\tilde{V}(1-\tilde{\phi}_m)}\right)^2+\left(\dfrac{2{\tilde{\phi}}_s}{\tilde{\eta}(1-2\tilde{\phi}_s)}\right)^2}}\right]-\tilde{\phi}_m=-\dfrac{2\kappa}{\tilde{R}}\left(1-\dfrac{A_0}{\tilde{A}}\right),
\label{pi_2}
\end{equation} 
where the quantities $\tilde{V}$, $\tilde{A}$ and $\tilde{R}$ now stand for the newly established internal volume, surface area and radius, respectively, of the stretched capsid after osmotic shock and $\tilde{\phi}_m=\frac{M}{\tilde{V}}$ is the new chain packing fraction. The strong increase in osmotic pressure resulting from conditions in which $\tilde{\phi}_s\ll \phi_s$ is assigned to the large solvent flux into the capsid, in an attempt to recover a situation of small contrast in ionic concentrations across the interface. The resulting size increase leads to swollen volumes $\tilde{V}$ and stretched areas $\tilde{A}$, which might eventually induce a mechanical instability against pore opening. Again, this analysis can be carried out based on Eq. (\ref{R_crit}), which dictates the maximum capsid size that ensures mechanical stability for a given set of elastic parameters. Together (\ref{phi_eq2}), (\ref{pi_1}), (\ref{pi_2}) and (\ref{R_crit}) provides us with the necessary theoretical tools to investigate capsid mechanical stability after and prior to osmotic shock. 

\section{Results}

We are now going to apply the theoretical framework described above to address the question of capsid stability under various external conditions. First, we shall consider some key features of the proposed model, such as some general trends of ionic absorption into the capsid and the resulting osmotic pressures under different ionic strengths. Next, we investigate how the osmotic equilibrium and the associated mechanical stability depend on the enclosed chain lengths and the buffer salt concentration. Finally, the question of mechanical stability against irreversible rupture driven by osmotic shock is investigated in detail.

\subsection{Osmotic equilibrium}

\begin{figure}[h!]
\includegraphics[width = 6.8cm,height = 4.5cm]{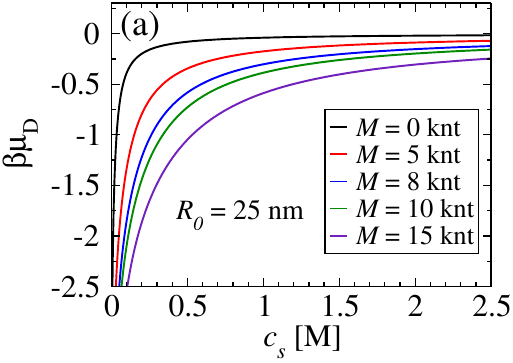}
\includegraphics[width = 6.8cm,height = 4.5cm]{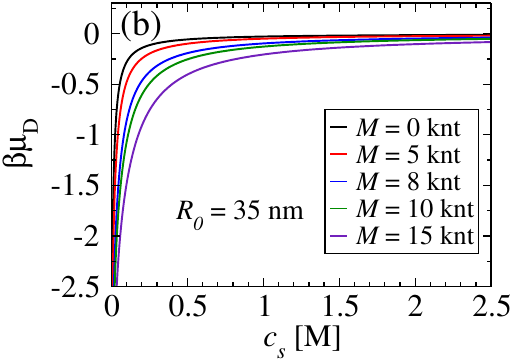}
\caption{Donnan potential across the capsid interface, as a function of the buffer ionic strength, in the case of packed genomes bearing different numbers of nucleotides, $M$. In (a), the capsid unstressed radius is $R_0=25$~nm, whereas in (b) a bigger size $R_0=35$~nm has been taken. In all cases, the capsid has a fixed surface charge of $\sigma_c=0.4e/\text{nm}^2$}.
\label{fig:fig3}
\end{figure}

\begin{figure}[h!]
\includegraphics[width = 7cm,height = 4.5cm]{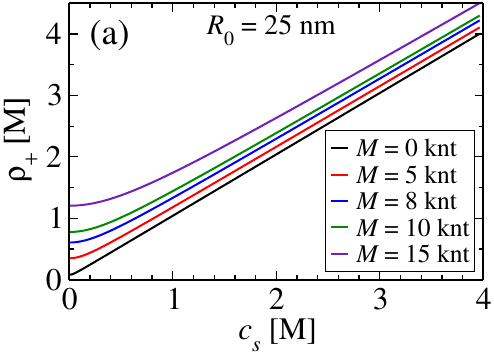}
\includegraphics[width = 7cm,height = 4.5cm]{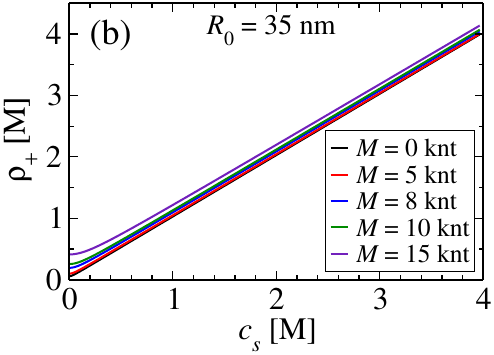}
\caption{Overall density of absorbed counterions as a function of the buffer salt concentration, considering different genome chain lengths $M$. The unstretched capsid sizes are $R_0=25$~nm (a) and $R_0=35$~nm (b), and the capsid surface charge is fixed at $\sigma_c=0.4e/\text{nm}^2$.}
\label{fig:fig4}
\end{figure}

We first analyze how the ionic diffusion and osmotic pressure across the capsid are influenced by the amount of added electrolyte in the outer solution. To this end, we consider that the ionic flow takes place at constant capsid volume. In this case, the equilibrium ionic concentrations depend crucially on the charge contrast between inner and outer solutions. Due to fixed charge of the packed genome, charge neutrality requires a large amount of counterions to be condensed into the capsid, along with those released by the charged capsid. This large counterion condensation, together with the requirement of equal chemical equilibrium across the interface, leads to a large potential drop across the surface. Obviously, size effects play a major role in such a Donnan equilibrium mechanism~\cite{Tin11}. All these trends can be observed in Fig.~\ref{fig:fig3}, which shows the Donnan potential, given by (\ref{mu_D2}), as a function of the external ionic strengths,  at several values of chain lengths $M$ typical of single strained genome, for the two representative capsid sizes $R=25$~nm (a) and $R=35$~nm (b). In all cases, a considerable potential difference is established across the interface at low ionic strengths, which becomes larger as $M$ increases. This leads to stronger Donnan potential for smaller capsid size observed in Fig.~\ref{fig:fig3}. Since the charge neutrality is enforced on a local level, the electrostatic energy is always zero. Therefore, the potential difference has in this case an indirect effect on the equilibrium balance, resulting in a strong entropic penalty for keeping a large concentration difference (to ensure electroneutrality) across the permeable membrane. This large contrast is made clear in Fig.~\ref{fig:fig4}, which shows the corresponding amount of absorbed counterions $N_{+}^0$ for the two situations depicted in Fig.~\ref{fig:fig3}. In both cases, the counterion concentration within the capsid quickly changes from a regime where $\rho_{+}\ll c_s$ at small buffer concentrations to a limiting regime $\rho_{+}\approx c_s$, in which the contrast between ionic concentrations becomes rather small. We also note that, in the case of a larger capsid, the effect of increasing $M$ is less pronounced since changing this quantity within the considered range does not lead to a considerable increase in the confined \textit{charge density}. We also note that the Donnan equilibrium in the proposed lattice model also incorporates exclusion volume effects.

\begin{figure}[h!]
\includegraphics[width=6.5cm,height=5cm]{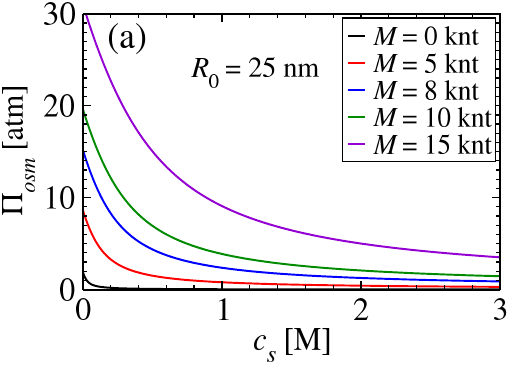}
\includegraphics[width=6.5cm,height=5cm]{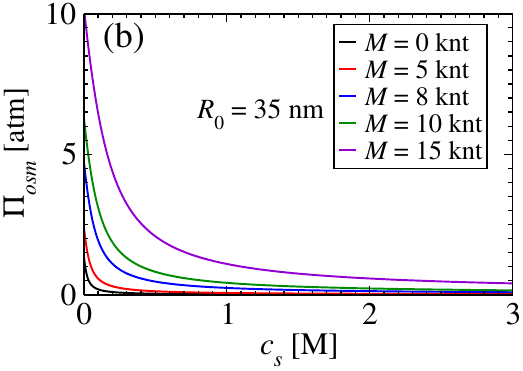}
\caption{Osmotic pressures, as predicted from Eq. (\ref{pi_1}), corresponding to the two representative cases under investigation. Again, the unstretched capsid sizes are $R_0=25$~nm (a) and $R_0=35$~nm (b), and the surface charge is $\sigma_c=0.4e/\text{nm}^2$}.
\label{fig:fig5}
\end{figure}

The large difference observed in counterion concentrations across the interface brings about a large outward osmotic pressure that attempts to expand the surface. In order to illustrate this effect, the osmotic pressures from (\ref{pi_1}), corresponding to the cases addressed in Fig.~\ref{fig:fig3}, are shown in Fig.~\ref{fig:fig5}. These results clearly indicate a strong relevance of salt content on the mechanical stability of the embedded capsids. Apart from the counterion entropic effects, the higher pressures in the case of longer genome chains also reflect the entropy cost for packing the chain into coil-like conformations resulting from confinement effects. These chain contributions become clearer in the limit of large ionic concentrations, where effects from counterion absorption become rather small. In this regime, the osmotic pressures in Fig.~\ref{fig:fig5} attain approximately constant values, which can be ascribed to chain confinement contributions alone. It is important to stress that these contributions in the case of ideal self-avoiding chains are of purely entropic nature.

\subsection{Mechanical stability}

\begin{figure}[h!]
\includegraphics[width=6.5cm,height=4.5cm]{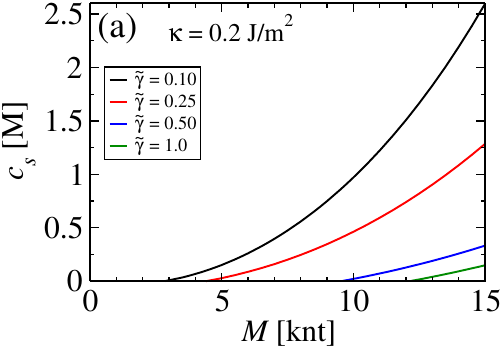}
\includegraphics[width=6.5cm,height=4.5cm]{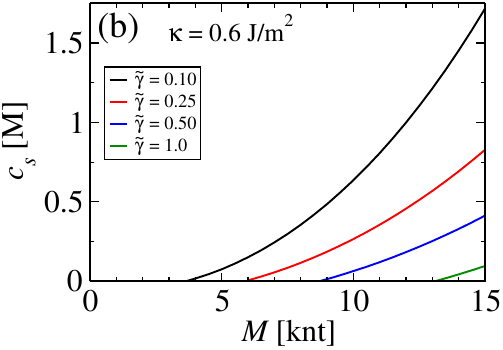}
\caption{Stability lines in the $(M,c_s)$ parameter plane for different elastic line tensions, and elastic modulus of $\kappa=0.2$~J/m$^2$ (a) and $\kappa=0.8$~J/m$^2$ (b). In each case, regions in parameter space lying to the left of the transition lines are stable ones, whereas points located to the right of the lines are unstable against pore formation.  }
\label{fig:fig6}
\end{figure}

The strong outward pressures resulting from osmotic equilibrium and chain confinement result in capsid stretching, at the cost of a surface elastic energy. If this expansion is sufficiently large, pore opening might become favorable. We shall here consider that, once this situation is achieved, the capsid looses its mechanical stability. This simple criteria allows us to map the regions in parameter space where the capsids will be certainly stable against pore nucleation. For a given set of elastic parameters and unstressed sizes, ($\kappa,\gamma,R_0$), the critical stretching size $R_{crit}$ after pore formation can take place is directly calculated from (\ref{R_crit}). The osmotic and equilibrium conditions (\ref{phi_eq2}) and (\ref{pi_1}) are numerically solved to compute the system parameters (\textit{e. g.}, salt concentration and chain length) in which the threshold capsid value  $R_{crit}$ is achieved. Stability lines delimiting the stable regions can be drawn for several sets of intrinsic elastic parameters. This is done in Fig.~\ref{fig:fig6}, which shows transition stability lines in the ($M$,$c_s$) plane for two representative elastic modulus $\kappa=0.2$~J/m$^2$ (\ref{fig:fig6}a) and $\kappa=0.8$~J/m$^2$ (\ref{fig:fig6}b), for capsids with different reduced line tensions $\tilde{\gamma}=\beta\gamma\lambda_B^{-1}$. Here, $\lambda_B=\beta e^2/\varepsilon =0.72$~nm is the so-called Bjerrum length, which represents a typical distance between condensed counterions and an oppositely charged monomer. Following the chosen stability criteria, regions lying bellow the transition lines represent points in which mechanical stability can not be assured. The remaining regions are the ones in which mechanical stability is satisfied. For a given set of elastic parameters, there is a limiting chain length $M$ where the capsid is always stable when carrying shorter genomes. These limiting chain lengths are the ones in which  the transition lines cross the horizontal axis in Fig.~\ref{fig:fig6}. For small line tensions, the energy cost for breaking up surface bonds is relatively small, and the capsid is not able to sustain the strong osmotic pressures at large $M$ and small ionic concentrations $c_s$. As a consequence, the stability region is shrank towards high salt concentrations as $\tilde{\gamma}$ is reduced. As this parameter increases, the capsid becomes robust against pore opening, being able to withstand large stresses. At the smaller bound strength considered, $\tilde{\gamma}=0.1$, the capsids are unstable only over a small region corresponding to very small salt concentrations and large chains $M\approx 12$ knt.   As this parameter is increased to $\tilde{\gamma}=1.0$, the capsids become more susceptible to break upon stretching, and the larger chain size of $M=15$~knt considered is only stable at very high salt concentrations.

\begin{figure}[h!]
\includegraphics[width=6cm,height=4.5cm]{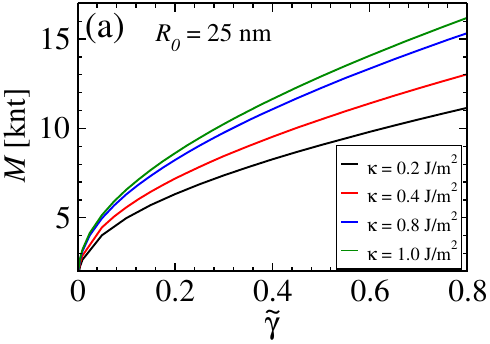}
\includegraphics[width=6cm,height=4.5cm]{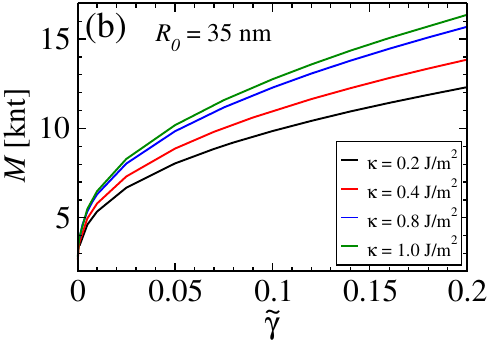}
\caption{Stability lines in the $(M,\tilde{\gamma})$ plane for different elastic modulus $\kappa$, for the case of capsids of undeformed sizes $R_0=25$~nm (a) and $R_0=35$~nm (b). In all cases, the capsids have a surface charge of $\sigma=0.4 e$/nm$^2$, and are in equilibrium with a buffer solution with a physiological salt concentration of $c_s=150$~mM. In these cases, points located to the right-side of the lines are stable ones, while those on the left represent unstable capsids.} 
\label{fig:fig7}
\end{figure}

In order to provide an overview on how elastic properties influence capsid stability for a given set of external parameters, we now consider the case of capsids within a wider range of elastic parameters, embedded on a buffer solution of physiological salt concentration, $c_s=150$~mM. Fig.~\ref{fig:fig7} shows the stability lines in the ($M,\tilde{\gamma}$), at several different elastic modulus $\kappa$, considering capsids of  unstressed sizes $R_0=25$~nm (\ref{fig:fig7}a) and $R_0=35$~nm (\ref{fig:fig7}b), for chain lengths lying in the range $0<M<15$~knt. Note that the stable points are now the ones lying below the transition lines. When the line tension $\tilde{\gamma}$ is sufficiently small, the capsids are very sensitive to small surface strains, such that the stable chain lengths do not depend strongly on the surface elasticity. We note, however, that the bigger capsid is able to enclose longer genome at much smaller binding strengths $\tilde{\gamma}$ (note the different scales in the horizontal axis of Fig.~\ref{fig:fig7}). This is clearly a result of the corresponding weaker pressures, as both the charge density and packing fraction of the encapsulated genome are smaller in this case. As $\tilde{\gamma}$ grows further, the surface rigidity starts to play a relevant role, as the surface is able to endure larger deformations without loosing its mechanical stability. The less deformable the capsids, the larger the genome size it is able to carry without loosing its integrity -- a trend which becomes more pronounced at larger $\tilde{\gamma}$, as shown in Fig. \ref{fig:fig7}. 

\subsection{Osmotic shock}

\begin{figure}[h!]
\includegraphics[width=6.5cm,height=4.5cm]{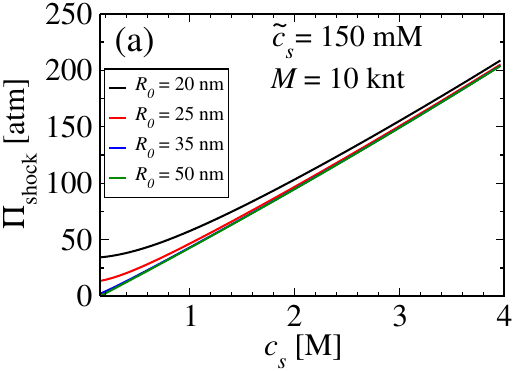}
\includegraphics[width=6.5cm,height=4.5cm]{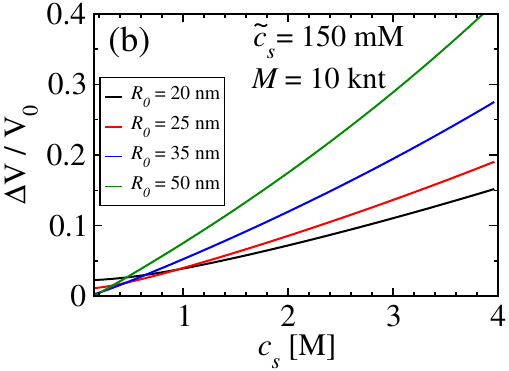}
\caption{Effects of osmotic shock on the mechanical equilibrium of capsids of different sizes, when the initial salt concentration $c_s$ (horizontal axis) is reduced to $\tilde{c}_s=150$~mM. In (a), the osmotic pressures after shock are shown as functions of the initial salt concentration, whereas panel (b) shows the relative swelling resulting from these pressures. In all cases, the capsids have a surface charge of $\sigma_c=0.4e$/nm$^2$ and surface strength of $\kappa=0.2$~J/m$^2$, while the packed genome has a length of $M=10$~knt.} 
\label{fig:fig8}
\end{figure}

So far, we have considered osmotic and mechanical equilibrium with an external solution of fixed concentration. We now turn our attention to a second situation, in which the ionic strength at the buffer solution is promptly reduced, after such equilibrium condition has been achieved. As stressed before, the new equilibrium condition established shortly after salt removal is given by (\ref{pi_2}). The high contrast in salt concentration results in a strong outward pressure at the capsid inner wall, which in turn is reflected as a large solvent intake that swells the capsid. These effects are summarized in Fig.~\ref{fig:fig8}, which shows the osmotic pressures (left panel) and the resulting volume increase (right panel) of capsids of various sizes carrying a genome $M=10$~knt long, driven by a quick dilution in buffer salt concentration from $c_s$ to a smaller, physiological salt concentration $\tilde{c}_s=150$~mM. The degree of particle swelling is a rather important quantity, since this measurement allows one to relate capsid rupture to its expansion either in virus maturation processes or shortly before capsid disassembly~\cite{Anv18}.

\begin{figure}[h!]
\includegraphics[width=5.7cm,height=4cm]{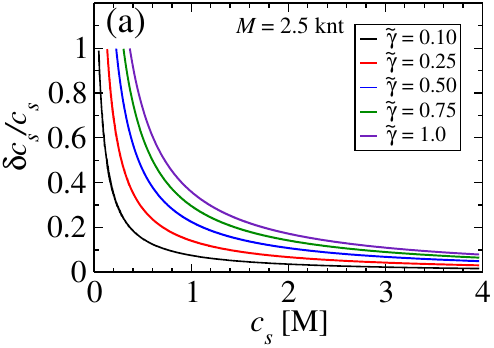}
\includegraphics[width=5.7cm,height=4cm]{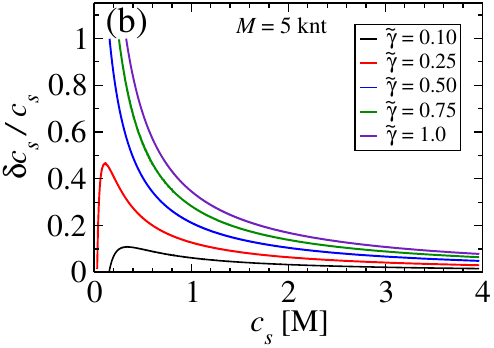}\\
\includegraphics[width=5.7cm,height=4cm]{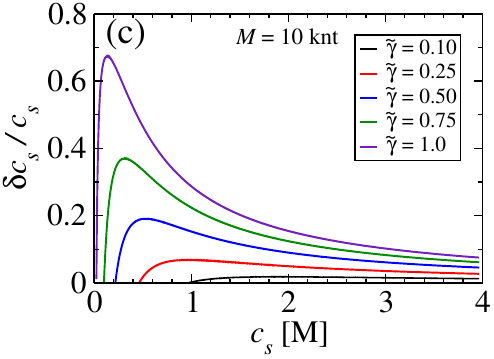}
\includegraphics[width=5.7cm,height=4cm]{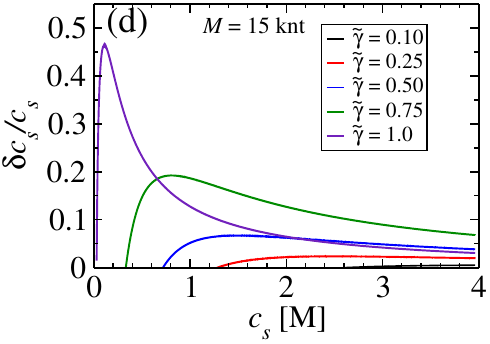}
\caption{Relative amount of salt that has to be removed from the external environment in order to induce capsid rupture \textit{via} osmotic shock, as a function of the initial ionic concentration $c_s$ for various chain lengths $M$ and binding strengths $\tilde{\gamma}$. The capsids have a unstressed size of $R_0=25$~nm, surface strength of $\kappa=0.2$~J/m$^2$, and a fixed charge density of $\sigma=0.4e/$nm$^{2}$.}
\label{fig:fig9}
\end{figure}

In real situations, many of the over-sized capsids predicted by overall mechanical equilibrium in Fig. \ref{fig:fig8}b will hardly be able to sustain such large surface inflation, and capsid rupture is very likely to take place. To address this point,  we now proceed to investigate the conditions that ensure the validity of mechanical stability, within a reasonably large range of initial salt conditions. This analysis can be performed using condition (\ref{pi_2}) to determine the extent of salt dilution $\delta c_s=c_s-\tilde{c}_s$ in the capsid's surroundings for which the capsid will be stretched to its critical size $R_{crit}$. Using this condition, we have computed the relative dilution $\delta c_s/c_s$ in the buffer solution necessary to drive capsid instability, as a function of its initial concentration $c_s$. The analysis is carried out over regions in ionic concentrations where the capsid is initially stable against rupture (even though it is always stretched, to some extend, in this initial condition). The results are shown in Fig.~\ref{fig:fig9}, where the required relative dilutions $\delta c_s/c_s$ to induce pore opening are shown as a function of the initial ionic concentration at several capsid strengths $\tilde{\gamma}$ and at different genome lengths $M$, for a capsid of fixed radius $R_0=25$~nm. 

An interesting behavior is observed in all cases, where the fraction of extracted salt crosses over from a strict monotonic behavior at the smallest chain length considered, $M=2.5$~knt (Fig.~\ref{fig:fig9}a), to a non-monotonic behavior at all $\tilde{\gamma}$ for the longest chains comprised of $M=10$~knt (Fig.~\ref{fig:fig9}c) and $M=15$~knt (Fig.~\ref{fig:fig9}d). The reason for these marked qualitative differences can be understood in terms of the initial conditions of capsid stress and the amount of adsorbed ions. If the genome packing fraction is sufficiently small, the amount of condensed counterions into the capsid is not very high, and a small dilution in salt concentration at the buffer solution does not lead to a significant contrast in ionic concentrations across the interface. As a result, the established outward pressures upon a small amount of salt removal are not high enough to drive capsid rupture. The capsid is then stable against rupture driven by osmotic-shock, whatever the amount of external ionic dilution. This situation lasts until a point where the capsid becomes unstable upon removal of the whole amount of external salt (corresponding to $\delta c_s/c_s=1$). After this point is reached, increasing of the initial salt concentration always results in mechanical instability after a given amount of salt dilution.

The above scenario starts to change at the intermediate chain length $M=5$~knt (Fig.~\ref{fig:fig9}b). For large binding strengths $\tilde{\gamma}$, the capsids are initially stable over the whole range of ionic concentrations. Therefore, there is a region of small salt concentrations where the capsids remain stable against osmotic shock, even if the whole content of external salt is removed. As the initial salt concentration increases, a critical value is reached in which the capsid becomes unstable upon full dilution of external salt. Further increase of $c_s$ beyond this point will require an increasingly small amount of salt dilution in order to destabilize the capsid \textit{via} osmotic shock. On the other hand, if $\tilde{\gamma}$ is small enough, the capsid will be initially unstable at small salt concentrations. At the smallest salt concentration $c_s$ beyond which the capsid becomes stable, the capsid is at critical size, $R_{crit}$. Now, as the initial salt concentration increases, the capsid becomes more stable against rupture before osmotic shock. A larger dilution is then necessary to trigger capsid instability \textit{via} osmotic shock, such that the amount of extracted salt $\delta c_s$ initially increases. As the chain length is increased (see Figs.~\ref{fig:fig9}c and \ref{fig:fig9}d), the capsids are always unstable at small salt concentrations for all values of $\tilde{\gamma}$ under investigation, thus suppressing the strict monotonic decay in $\delta c_s$ that leads to capsid rupture. We note this non-trivial behavior can be simply understood in terms of the role played by ionic adsorption on the osmotic pressure across the capsids.

\section{Conclusion}

We presented a simple theoretical description, based on a combined ion-chain lattice gas model, to address the question of capsid osmotic and mechanical stability and their robustness against changes on the external surroundings. The proposed approach is quite general, and therefore applicable to a wide class of system parameters. As usual, the main drawback of such generally-aimed approaches in the description of physical systems displaying a large variety of sizes, shapes, mechanical properties and internal composition is the intrinsic dependence on a number of  coarse-graining parameters, which are to be further matched to describe specific system conditions. Still, the proposed lattice model is able to underpin some of the key features dictating mechanical stability and its dependency on external conditions. Examples of such relevant physical contributions include the exclusion volume effects upon strong confinement, the entropy reduction due to the limited allowed conformations inherent to the confined self-avoiding chain, as well the electroneutrality condition and the underlying ionic entropy loss. All these mechanisms  are partially accounted for in the proposed lattice-gas formalism.

A great advantage of the proposed theoretical description lies on its ease of implementation. With virtually no computation cost, a large range of system parameters can be explored to identify regions where mechanical stability holds. We have then used the model to predict the stability regions in parameter space (\textit{i. e.}, those in which pore nucleation is energetically unfavorable), considering the situations of fixed elastic parameters, as well as situations of varying monomer lengths and a fixed, physiological salt concentration. Since the model is based on a non-local approach based on the mean concentrations of absorbed ions, it does not account for the effects of large inhomogeneities. Improvements of the proposed model to properly describe these situations require the incorporation of  non-local effects, which can be accomplished {\textit via} extensions along the lines of Poisson-Boltzmman approach and its modified versions that account for finite size effects. Moreover, effects from chain architecture (\textit{e. g.}, quenched/annealed branching) and conformation can be incorporated into the Flory description~\cite{Scho13} to provide further insights on the role played by the packed ssRNA on mechanical stability.

\section{Acknowledgments}

This work was partially supported by the PIBIC-CNPq program.

\appendix\section{Free energies of a chain-ion mixture in the lattice-model framework}

We now present in detail the statistical calculations that lead to the lattice-gas free-energies. The free energy of the non-interacting lattice system only comprises the contributions from mixing entropy, related to the distinct allowed configuration for the multi-component system within the lattice matrix. Such exclusion volume contributions lead to significant entropy losses with respect to a point-like particle description, and become very pronounced (also playing a major role) in the case of strongly confined mixtures. The entropy of multi-component system is computed via
\begin{equation}
S=k_B\ln(\Omega),
\label{S_t1}
\end{equation}
where, as usual, $\Omega$ denotes the number of allowed particle configurations. To calculate this quantity, we first consider the possible ways of assembling the fixed $M$ monomer beads inside $V$ lattice sites, where $V b^3$ represent the capsid volume, $b$ being a typical size-scale of a single lattice site. We notice in passing that the number of such configurations is closely related to the possible dynamic pathways displayed by a sequence of $M$ connected lattice-sites -- the so-called \textit{Self-Avoiding Walk} (SAW). Let $\Omega_{chain}(M,V)$ be the number of configurations for a homopolymer chain of length $M$ confined in the lattice volume $V$. For each such configuration, there are $\Omega_{ion}(M,V,N_+,N_-)$ ways of arranging the mobile ions in the remaining empty sites. The total number of configurations is thus $\Omega=\Omega_{chain}\Omega_{ion}$, such that the entropic mixing contribution in (\ref{S_t1}) decouples into chain and ionic contributions,
\begin{equation}
S=k_B\ln(\Omega_{chain}\Omega_{ion})=k_B\ln(\Omega_{chain})+k_B\ln(\Omega_{ion})\equiv S_{chain}+S_{ion}.
\label{S_t2}
\end{equation}
 In order to calculate $S_{chain}$, we employ here a mean-field approach in which the $M$ chain beads are allocated in sequence across neighboring lattice sites. There are $V$ possible ways of allocating the first bead in an arbitrary lattice site, such that $\Omega_m(j=1,V)=V$ is the corresponding number of states. Since there are already occupied sites in the lattice, addition of subsequent beads must account for the probability that a neighboring cell is empty, such that 
 \begin{equation}
 \Omega(j+1,V)=\Omega(j,V)P(j,V)(\ell-1),
 \label{Om_chain0}
 \end{equation}
where $\ell$ is the lattice coordination number, and $P(j,V)$ stands for the probability of finding an empty site, given that $j$ sequential sites have been occupied by the previous polymer beads. In general, this quantity is non-local, and depends both on boundary effects and the particular configuration in which the previous $j$ beads have been arranged over the lattice. However a simple and physically transparent description can be used by invoking a Flory mean-field approach, whereby the probability to $P(j,V)$ is taken to be non-local, corresponding to the overall probability of randomly select one of the $V-j$ unoccupied ones out of $V$ ones, $P(j,V)=(V-j)/V$. Once the first bead has been placed, the number of configurations for a two-bead chain is thus $\Omega(2,V)=\Omega(1,V)(\ell-1)V/(V-1)=(\ell-1)V(V-1)/V$, that of a three-bead chain is $\Omega(3,V)=\Omega(2,V)(\ell-1)(V-2)/V=(\ell-1)^2V(V-1)(V-2)/V^2$, and so on. Proceeding along these lines all over the chain length, up to its last $M$-th bead, we find
\begin{equation}
\begin{aligned}
\Omega_{chain}(M,V)  =\prod_{j=1}^{M}\Omega(j,V)P(j,V)& =V(\ell-1)\left(\dfrac{V-1}{V}\right)(\ell-1)\left(\dfrac{V-2}{V}\right)(\ell-1)\left(\dfrac{V-3}{V}\right)\\
&\cdots(\ell-1)\left(\dfrac{V-M+1}{V}\right)=\dfrac{(\ell-1)^{M-1}V!}{V^{M-1}(V-M)!}.
\end{aligned}
\label{Om_chain1}
\end{equation}     
Taking the logarithm of both sides of this expression, and further invoking the Stirling's approximation for $V\gg 1$ and $M\gg 1$ (in such a way that $\phi_m=M/V$ remains constant), one finds
\begin{equation}
\ln(\Omega_{chain}(M))\equiv \dfrac{S_{chain}}{k_B}=M[\ln(\ell-1)-1]+(M-V)\ln\left(\dfrac{V-M}{V}\right)+\ln(V),
\label{Om_chain2}
\end{equation}  
In the limit $V\gg 1$, the last term on the r.h.s can be neglected with respect to the linear term in $V$. Therefore, in the thermodynamic limit the entropic chain contribution reads 
\begin{equation}
\dfrac{S_{chain}}{k_B}=M[\ln(\ell-1)-1]+(M-V)\ln\left(1-\phi_m\right).
\label{Schain}
\end{equation}
Now, for each $\Omega(V,M)$ chain configuration, there are $\Omega_{ion}$ possible ways of arranging $N_{+}$ cations and $N_{-}$ anions among the remaining empty sites. Since $M$ out of $V$ sites are already occupied by the fixed chain monomers, the number of allowed ionic configurations corresponds to the possible ways of arranging $N_{+}$ and $N_{-}$ distinct objects into $V-M$  empty boxes, which is given by 
\begin{equation}
\Omega_{ion}(M,V,N_{+},N_{-})=\dfrac{(V-M)!}{(V-M-N_{+}-N_{-})!N_{+}!N_{-}!}
\label{Om_ion}
\end{equation}
Taking the logarithm and once again invoking Stirling's approximation, one finds
\begin{equation}
\begin{aligned}
\dfrac{S_{ion}}{k_B}=k_B\biggr[(M-V)\ln\left(\dfrac{V-M-N_+-N_-}{V-M}\right)+N_+\ln\left(\dfrac{V-M-N_+-N_-}{N_+}\right)\\
+N_-\ln\left(\dfrac{V-M-N_+-N_-}{N_-}\right)\biggr].
\end{aligned}
\label{S_ion1}
\end{equation}
After some algebraic manipulations, the relation above can be rewritten as
\begin{equation}
\begin{aligned}
\dfrac{S_{ion}}{k_B}=\biggr[(M-V+N_{+}+N_{-})\ln(1-\phi_m-\phi_{-}-\phi_{+})-N_{+}\ln(\phi_{+})-N_{-}\ln(\phi_{-})\\
-(V-M)\ln(1-\phi_m)\biggr].
\end{aligned}
\label{S_ion2}
\end{equation}
Now, the corresponding free internal energy can be computed from the definition $\mathcal{F}_{in}=U-TS$. In the absence of direct site-site interactions, $U=0$, and we are left with
\begin{equation}
\mathcal{F}_{in}=-TS\rightarrow \beta\mathcal{F}=-\dfrac{S_{chain}}{k_B}-\dfrac{S_{ion}}{k_B}\equiv \beta\mathcal{F}_{chain}+\beta\mathcal{F}_{ion},
\label{F_inap1}
\end{equation}
where we have defined the chain and ionic free energies as $\beta\mathcal{F}_{chain}=-S_{chain}/k_B$ and $\beta\mathcal{F}_{ion}=-S_{ion}/k_B$, respectively. Finally, combination of these definitions with (\ref{Schain}) and (\ref{S_ion2}) leads to (\ref{fc}) and (\ref{f_ions}).
As for the external free energy contributions, an expression similar to (\ref{Om_ion}) can be obtained for the total amount of allowed configurations, where now the available volume $V-M$ is replaced by $V'$:
 \begin{equation}
\Omega'_{ion}(V',N'_{+},N'_{-})=\dfrac{V'!}{(V'-N'_{+}-N'_{-})!N'_{+}!N'_{-}!},
\label{Omp_ion}
\end{equation}
where $\Omega'$ stands for the number of configurations in the outer buffer. The corresponding mixing entropy reads as 
 \begin{equation}
\dfrac{S_{out}}{k_B}=(N_{+}'+N_{-}'-V)\ln(1-\phi_{+}'-\phi_{-}')-N_{-}'\ln(\phi'_{-})-N'_{+}\ln(\phi'_{+}),
\label{Omp_ion}
\end{equation}
which leads directly to the mixing free energy in the external solution, (\ref{f_ex1}). 

\section{Classical nucleation theory of pore formation}

\begin{figure}[h!]
\centering
\includegraphics[height = 5cm, width = 6cm]{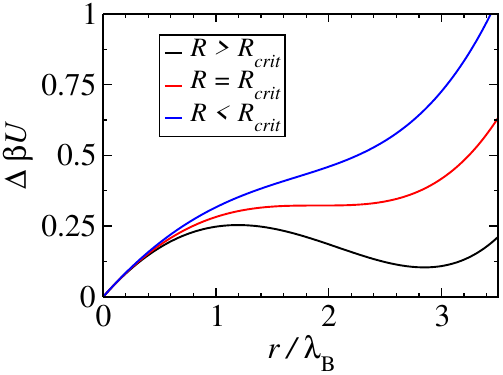}
\caption{Energy profiles as functions of pore sizes, showing the disappearance of a nucleation barrier as $R$ increases beyond the critical size.}
\label{fig:fig10}
\end{figure}

We now discuss some key aspects of the employed nucleation-like theory for pore formation, which is modeled as a competition between a surface energy gain and a line energy penalty upon pore opening. In a mean-level description, these contributions are  described, respectively, by the first and second terms on the r.h.s of (\ref{del_U}), which considers the opening of a circular surface hole of radius $r$, such that the pore surface area is $A_p=\pi r^2$. In terms of this quantity (\ref{del_U}) reads as
\begin{equation}
\Delta U(r)=\dfrac{\kappa}{2A_0}\left[A_p^2-2A_p(A-A_0)\right]+2\pi r\gamma.
\label{del_U2}
\end{equation}
This relation can be further written in terms of pore and surface radii as
\begin{equation}
\Delta U(r)=\dfrac{\pi\kappa}{8\pi R_0^2}\left[r^4-2r^2(R^2-R_0^2)\right]+2\pi r\gamma.
\label{del_U3}
\end{equation}
At very small pore radius, the last term of this relation dominates, and the energy cost for pore opening grows linearly with pore radius. As the pore size increases, the first term becomes dominant, leading to an unbound growth of energy penalty at larger $r$. Depending on the particular set of system parameters, an intermediate regime might emerge in which the influence of the second term leads to a decrease in energy cost upon increasing in pore size. In this case, $\Delta U(r)$ displays a second minimum beyond the one at vanishing pore size. The minimum appears after a local maximum, whose height is usually referred to as the \textit{nucleation barrier}. If a thermal fluctuation is large enough to overcome such an energy barrier, the system is allowed to ``jump'' into a transient state represented by the second local minimum. At this point, a new dynamical equilibrium will be settled in which the pore can either close-up again or grow further, leading to irreversible shell rupture. If the surface is weakly stressed, $R\sim R_0$, the function $\Delta U(r)$ is a monotonically increasing function of $r$, and the capsid is stable against pore formation. There is a threshold surface radius $R_{crit}$ corresponding to the onset of the second energy stable point, thereby delimiting the two mechanical stability regimes. All these features are illustrated in Fig. \ref{fig:fig10}. 

The stationary points of (\ref{del_U3}) are the ones satisfying $\Delta U'(r)=0$, from which the following algebraic relation is obtained
\begin{equation}
\Delta U'(r)=r^3-4(R^2-R_0^2)r+\dfrac{4 \gamma R_0^2}{\kappa}=0.
\label{del_U3}
\end{equation}
This is a cubic equation of depressed type, $r^3+pr+q=0$. Since $U(r=0)=q>0$, there is necessarily one real, negative root, to this equation. The two other roots are either a complex conjugate pair or two real-valued numbers. In the latter case, the two remaining roots are positive, since $p<0$ implies $\Delta U'(r=0)<0$ (recall that $R>R_0$ for the stretched capsid). These values correspond to the pore sizes in which $\Delta U(r)$ achieves their local extrema in Fig.~\ref{fig:fig9}. The three real roots to Eq (\ref{del_U3}) can be computed by the formula
\begin{equation}
r_m=\sqrt{\dfrac{-4p}{3}}\cos\left[\dfrac{\varphi-2\pi m}{3}\right],
\label{roots}
\end{equation}
where $\varphi\equiv\cos^{-1}\left(\dfrac{3q}{2p}\sqrt{\dfrac{-3}{p}}\right)$ and $m=0,1,2$. The requirement for (\ref{del_U3}) to have these three real roots (in which case $\Delta U(r)$ features two  stationary points that might lead to irreversible pore nucleation) is that $4p^3+27q^2<0$. Conversely, if $4p^3+27q^2\le 0$, the function  $\Delta U(r)$ increases monotonically at all pore sizes, which is a sufficient (though not necessary) condition that guarantees mechanical stability against pore formation. For a given set of elastic parameters and unstressed shell size $R_0$, there will be a critical stretched surface size $R_{crit}$ beyond which mechanical integrity might be compromised. This threshold value is determined by the condition  $4p^3+27q^2=0$, which in terms of the underlying  system parameters reads as
\begin{equation}
(R_{crit}^2-R_0^2)=\left(\dfrac{3}{4}\right)^3\left(\dfrac{2\gamma R_0^2}{\kappa}\right)^2
\label{roots}
\end{equation}
which, after some algebraic manipulations, leads directly to the condition (\ref{R_crit}) for the critical capsid size able to ensure mechanical stability.


\begin{thebibliography}{10}

\bibitem{Cas62}
D.~L. Caspar and A.~Klug, ``Physical principles in the construction of regular
  viruses,'' {\em Cold Spring Harbor symposia on quantitative biology},
  vol.~27, pp.~1--24, 1962.

\bibitem{Tre06}
T.~Douglas and M.~Young, ``Viruses: Making friends with old foes,'' {\em
  Science}, vol.~312, no.~5775, pp.~873--875, 2006.

\bibitem{Ros89}
M.~G. Rossmann and J.~E. Johnson, ``Icosahedral rna virus structure,'' {\em
  Annual Review of Biochemistry}, vol.~58, no.~1, pp.~533--569, 1989.

\bibitem{Zlo04}
A.~Zlotnick, ``Viruses and the physics of soft condensed matter,'' {\em
  Proceedings of the National Academy of Sciences}, vol.~101, pp.~15549--15550,
  2004.

\bibitem{Roos10}
W.~H. Roos, R.~Bruinsma, and G.~J.~L. Wuite, ``Physical virology,'' {\em Nature
  Physics}, vol.~6, pp.~733--743, 2010.

\bibitem{Abr12}
N.~G. Abrescia, D.~H. Bamford, J.~M. Grimes, and D.~I. Stuart, ``Structure
  unifies the viral universe,'' {\em Annual Review of Biochemistry}, vol.~81,
  no.~1, pp.~795--822, 2012.

\bibitem{Mate13}
M.~G. Mateu, ``Assembly, stability and dynamics of virus capsids,'' {\em
  Archives of Biochemistry and Biophysics}, vol.~531, pp.~65 -- 79, 2013.

\bibitem{Luque2013}
A.~Luque and D.~Reguera, {\em Theoretical Studies on Assembly, Physical
  Stability and Dynamics of Viruses}, pp.~553--595.
\newblock Dordrecht: Springer Netherlands, 2013.

\bibitem{Zan20}
R.~Zandi, B.~Dragnea, A.~Travesset, and R.~Podgornik, ``On virus growth and
  form,'' {\em Physics Reports}, vol.~847, pp.~1--102, 2020.

\bibitem{Zhd21}
V.~P. Zhdanov, ``Virology from the perspective of theoretical colloid and
  interface science,'' {\em Current Opinion in Colloid \& Interface Science},
  vol.~53, p.~101450, 2021.

\bibitem{bach14}
M.~Bachmann, {\em Thermodynamics and Statistical Mechanics of Macromolecular
  Systems}.
\newblock Cambridge University Press, 2014.

\bibitem{Roo07}
W.~H. Roos, I.~L. Ivanovska, A.~Evilevitch, and G.~J. Wuite, ``Viral capsids:
  mechanical characteristics, genome packaging and delivery mechanisms,'' {\em
  Cellular and molecular life sciences : CMLS}, vol.~64, p.~1484–1497, 2007.

\bibitem{Bru15}
R.~F. Bruinsma and W.~S. Klug, ``Physics of viral shells,'' {\em Annual Review
  of Condensed Matter Physics}, vol.~6, no.~1, pp.~245--268, 2015.

\bibitem{Che18}
M.~Chevreuil, D.~Law-Hine, J.~Chen, S.~Bressanelli, S.~Combet, D.~Constantin,
  J.~Degrouard, J.~M\"oller, M.~Zeghal, and G.~Tresset, ``Nonequilibrium
  self-assembly dynamics of icosahedral viral capsids packaging genome or
  polyelectrolyte,'' {\em Nat. Commun.}, vol.~9, p.~3071, 2018.

\bibitem{Ree19}
R.~F. Garmann, A.~M. Goldfain, and V.~N. Manoharan, ``Measurements of the
  self-assembly kinetics of individual viral capsids around their rna genome,''
  {\em Proceedings of the National Academy of Sciences}, vol.~116, no.~45,
  pp.~22485--22490, 2019.

\bibitem{Ale03}
A.~Evilevitch, L.~Lavelle, C.~M. Knobler, E.~Raspaud, and W.~M. Gelbart,
  ``Osmotic pressure inhibition of dna ejection from phage,'' {\em Proceedings
  of the National Academy of Sciences}, vol.~100, no.~16, pp.~9292--9295, 2003.

\bibitem{Zan05}
R.~Zandi and D.~Reguera, ``Mechanical properties of viral capsids,'' {\em Phys.
  Rev. E}, vol.~72, p.~021917, 2005.

\bibitem{Evi11}
A.~Evilevitch, W.~H. Roos, I.~L. Ivanovska, M.~Jeembaeva, B.~Jönsson, and
  G.~J. Wuite, ``Effects of salts on internal dna pressure and mechanical
  properties of phage capsids,'' {\em Journal of Molecular Biology}, vol.~405,
  no.~1, pp.~18--23, 2011.

\bibitem{Jin12}
Z.~Jin and J.~Wu, ``Density functional theory for encapsidated
  polyelectrolytes: A comparison with monte carlo simulation,'' {\em The
  Journal of Chemical Physics}, vol.~137, no.~4, p.~044905, 2012.

\bibitem{Cor03}
A.~Cordova, M.~Deserno, W.~M. Gelbart, and A.~Ben-Shaul, ``Osmotic shock and
  the strength of viral capsids,'' {\em Biophysical Journal}, vol.~85,
  pp.~70--74, 2003.

\bibitem{Col20}
T.~Colla, A.~Bakhshandeh, and Y.~Levin, ``Osmotic stress and pore nucleation in
  charged biological nanoshells and capsids,'' {\em Soft Matter}, vol.~16,
  pp.~2390--2405, 2020.

\bibitem{Zen21}
C.~Zeng, L.~Scott, A.~Malyutin, R.~Zandi, P.~Van~der Schoot, and B.~Dragnea,
  ``Virus mechanics under molecular crowding,'' {\em The Journal of Physical
  Chemistry B}, vol.~125, no.~7, pp.~1790--1798, 2021.

\bibitem{Alz21}
M.~O. Alziyadi and A.~R. Denton, ``Osmotic pressure and swelling behavior of
  ionic microcapsules,'' {\em The Journal of Chemical Physics}, vol.~155,
  no.~21, p.~214904, 2021.

\bibitem{Suk21}
L.~Suken{'\i}k, L.~Mukhamedova, M.~Proch{\'a}zkov{\'a}, K.~\v{S}kubn{\'i}k,
  P.~Plevka, and R.~V{\'a}cha, ``Cargo release from nonenveloped viruses and
  virus-like nanoparticles: Capsid rupture or pore formation,'' {\em ACS Nano},
  vol.~15, no.~12, pp.~19233--19243, 2021.

\bibitem{Almendral}
J.~M. Almendral, ``Assembly of simple icosahedral viruses,'' in {\em Structure
  and Physics of Viruses} (M.~Mateu, ed.), pp.~307--328, Springer, Dordrecht,
  2013.

\bibitem{Keg06}
W.~K. Kegel and P.~{van der Schoot}, ``Physical regulation of the self-assembly
  of tobacco mosaic virus coat protein,'' {\em Biophysical Journal}, vol.~91,
  no.~4, pp.~1501--1512, 2006.

\bibitem{Kat10}
S.~P. Katen, S.~R. Chirapu, M.~G. Finn, and A.~Zlotnick, ``Trapping of
  hepatitis b virus capsid assembly intermediates by phenylpropenamide assembly
  accelerators,'' {\em ACS Chemical Biology}, vol.~5, no.~12, pp.~1125--1136,
  2010.

\bibitem{Rij13}
P.~van Rijn, M.~Tutus, C.~Kathrein, L.~Zhu, M.~Wessling, U.~Schwaneberg, and
  A.~Böker, ``Challenges and advances in the field of self-assembled
  membranes,'' {\em Chem. Soc. Rev.}, vol.~42, pp.~6578--6592, 2013.

\bibitem{Mat13}
R.~Matthews and C.~N. Likos, ``Dynamics of self-assembly of model viral capsids
  in the presence of a fluctuating membrane,'' {\em The Journal of Physical
  Chemistry B}, vol.~117, no.~27, pp.~8283--8292, 2013.

\bibitem{Hagan14}
M.~F. Hagan, {\em Modeling Viral Capsid Assembly}, ch.~1, pp.~1--68.
\newblock John Wiley \& Sons, Ltd, 2014.

\bibitem{Per15}
J.~D. Perlmutter and M.~F. Hagan, ``Mechanisms of virus assembly,'' {\em Annual
  Review of Physical Chemistry}, vol.~66, no.~1, pp.~217--239, 2015.

\bibitem{Bru16}
R.~F. Bruinsma, M.~Comas-Garcia, R.~F. Garmann, and A.~Y. Grosberg,
  ``Equilibrium self-assembly of small rna viruses,'' {\em Phys. Rev. E},
  vol.~93, p.~032405, 2016.

\bibitem{Twa19}
R.~Twarock and A.~Luque, ``Structural puzzles in virology solved with an
  overarching icosahedral design principle,'' {\em Nat. Commun.}, vol.~10,
  p.~4414, 2019.

\bibitem{Men20}
C.~I. Mendoza and D.~Reguera, ``Shape selection and mis-assembly in viral
  capsid formation by elastic frustration,'' {\em eLife}, vol.~9, p.~e52525,
  2020.

\bibitem{Mil15}
P.~Miles, P.~Cassidy, L.~Donlon, O.~Yarkoni, and D.~Frankel, ``In vitro
  assembly of a viral envelope,'' {\em Soft Matter}, vol.~11, pp.~7722--7727,
  2015.

\bibitem{Jus20}
J.~Spiriti, J.~F. Conway, and D.~M. Zuckerman, ``Should virus capsids assemble
  perfectly? theory and observation of defects,'' {\em Biophysical Journal},
  vol.~119, no.~9, pp.~1781--1790, 2020.

\bibitem{Vla06}
V.~A. Belyi and M.~Muthukumar, ``Electrostatic origin of the genome packing in
  viruses,'' {\em Proceedings of the National Academy of Sciences}, vol.~103,
  no.~46, pp.~17174--17178, 2006.

\bibitem{Sil08_2}
A.~\ifmmode~\check{S}\else \v{S}\fi{}iber and R.~Podgornik, ``Nonspecific
  interactions in spontaneous assembly of empty versus functional
  single-stranded rna viruses,'' {\em Phys. Rev. E}, vol.~78, p.~051915, 2008.

\bibitem{Peng12}
P.~Ni, Z.~Wang, X.~Ma, N.~C. Das, P.~Sokol, W.~Chiu, B.~Dragnea, M.~Hagan, and
  C.~C. Kao, ``An examination of the electrostatic interactions between the
  n-terminal tail of the brome mosaic virus coat protein and encapsidated
  rnas,'' {\em Journal of Molecular Biology}, vol.~419, no.~5, pp.~284--300,
  2012.

\bibitem{Gar15}
R.~F. Garmann, M.~Comas-Garcia, C.~M. Knobler, and W.~M. Gelbart, ``Physical
  principles in the self-assembly of a simple spherical virus,'' {\em Accounts
  of Chemical Research}, vol.~49, no.~1, pp.~48--55, 2016.
\newblock PMID: 26653769.

\bibitem{Yang22}
Y.~Wang and T.~Douglas, ``Bioinspired approaches to self-assembly of virus-like
  particles: From molecules to materials,'' {\em Accounts of Chemical
  Research}, vol.~55, no.~10, pp.~1349--1359, 2022.

\bibitem{Tando16}
G.~Erdemci-Tandogan, J.~Wagner, P.~van~der Schoot, R.~Podgornik, and R.~Zandi,
  ``Effects of rna branching on the electrostatic stabilization of viruses,''
  {\em Phys. Rev. E}, vol.~94, p.~022408, Aug 2016.

\bibitem{Cas13}
C.~J. R. and C.~J.L., ``The basic architecture of viruses,'' in {\em
  Subcellular Biochemistry} (M.~Mateu, ed.), vol.~68, Springer, Dordrecht,
  2013.

\bibitem{Mar17}
S.~Marion, C.~{San Mart{\'i}n}, and A.~\v{S}iber, ``Role of condensing
  particles in polymer confinement: A model for virus-packed
  “minichromosomes”,'' {\em Biophysical Journal}, vol.~113, no.~8,
  pp.~1643--1653, 2017.

\bibitem{Spa05}
A.~J. Spakowitz and Z.-G. Wang, ``Dna packaging in bacteriophage: Is twist
  important?,'' {\em Biophysical Journal}, vol.~88, no.~6, pp.~3912--3923,
  2005.

\bibitem{Hir15}
A.~D. Hirsh and N.~Perkins, ``Dna buckling in bacteriophage cavities as a
  mechanism to aid virus assembly,'' {\em Journal of Structural Biology},
  vol.~189, no.~3, pp.~251--258, 2015.

\bibitem{Vet15}
R.~Vetter, F.~K. Wittel, and H.~J. Herrmann, ``Packing of elastic wires in
  flexible shells,'' {\em {EPL} (Europhysics Letters)}, vol.~112, no.~4,
  p.~44003, 2015.

\bibitem{Rap16}
D.~C. Rapaport, ``Packaging stiff polymers in small containers: A molecular
  dynamics study,'' {\em Phys. Rev. E}, vol.~94, p.~030401, 2016.

\bibitem{flory}
P.~J. Flory, {\em Statistical Mechanics of Chain Molecules}.
\newblock New York: John Wiley, 1969.

\bibitem{Lev02}
Y.~Levin, ``Electrostatic correlations: from plasma to biology,'' {\em Rep.
  Prog. Phys}, vol.~65, no.~11, p.~1577, 2002.

\bibitem{Lev96}
Y.~Levin, ``Theory of counterion association in rod-like polyelectrolytes,''
  {\em Europhys. Lett.}, vol.~34, no.~6, pp.~405--410, 1996.

\bibitem{Kuh98}
P.~S. Kuhn, Y.~Levin, and M.~C. Barbosa, ``Rodlike polyelectrolytes in the
  presence of monovalent salt,'' {\em Macromolecules}, vol.~31, no.~23,
  pp.~8347--8355, 1998.

\bibitem{Sil12}
A.~\v{S}iber, A.~L. Bo\v{z}i\v{c}, and R.~Podgornik, ``Energies and pressures
  in viruses: contribution of nonspecific electrostatic interactions,'' {\em
  Phys. Chem. Chem. Phys.}, vol.~14, pp.~3746--3765, 2012.

\bibitem{Tzli03}
S.~Tzlil, J.~T. Kindt, W.~M. Gelbart, and A.~Ben-Shaul, ``Forces and pressures
  in dna packaging and release from viral capsids,'' {\em Biophysical journal},
  vol.~84, no.~3, p.~1616–1627, 2003.

\bibitem{Pet07}
A.~S. Petrov and S.~C. Harvey, ``Structural and thermodynamic principles of
  viral packaging,'' {\em Structure}, vol.~15, no.~1, pp.~21--27, 2007.

\bibitem{Jeh15}
J.~Kim and J.~Wu, ``A thermodynamic model for genome packaging in hepatitis b
  virus,'' {\em Biophysical Journal}, vol.~109, no.~8, pp.~1689--1697, 2015.

\bibitem{Smi14}
G.~R. Smith, L.~Xie, B.~Lee, and R.~Schwartz, ``Applying molecular crowding
  models to simulations of virus capsid assembly in vitro,'' {\em Biophysical
  journal}, vol.~106, no.~1, p.~310–320, 2014.

\bibitem{Mee08}
M.~Jeembaeva, M.~Castelnovo, F.~Larsson, and A.~Evilevitch, ``Osmotic pressure:
  Resisting or promoting dna ejection from phage?,'' {\em Journal of Molecular
  Biology}, vol.~381, no.~2, pp.~310--323, 2008.

\bibitem{Bran19}
A.~Brandariz-Nu{\~n}ez, T.~Liu, T.~Du, and A.~Evilevitch, ``Pressure-driven
  release of viral genome into a host nucleus is a mechanism leading to herpes
  infection,'' {\em eLife}, vol.~8, p.~e47212, 2019.

\bibitem{Ian01}
I.~G. Macara, ``Transport into and out of the nucleus,'' {\em Microbiology and
  Molecular Biology Reviews}, vol.~65, no.~4, pp.~570--594, 2001.

\bibitem{Evi08}
A.~Evilevitch, L.~T. Fang, A.~M. Yoffe, M.~Castelnovo, D.~C. Rau, V.~A.
  Parsegian, W.~M. Gelbart, and C.~M. Knobler, ``Effects of salt concentrations
  and bending energy on the extent of ejection of phage genomes,'' {\em
  Biophysical Journal}, vol.~94, no.~3, pp.~1110--1120, 2008.

\bibitem{Jon11}
E.~Jo\'nczyk-Matysiak, M.~K\l{}ak, R.~Mi\k{e}dzybrodzki, and G.~Andrzej, ``The
  influence of external factors on bacteriophages—review,'' {\em Folia
  microbiologica}, vol.~56, pp.~191--200, 05 2011.

\bibitem{Chen12}
W.~Chen, H.~Zhang, L.~Gu, F.~Li, and F.~Yang, ``Effects of high salinity, high
  temperature and ph on capsid structure of white spot syndrome virus,'' {\em
  Diseases of aquatic organisms}, vol.~101, pp.~167--71, 11 2012.

\bibitem{Liu21}
P.~Liu, J.~Arsuaga, M.~C. Calderer, D.~Golovaty, M.~Vazquez, and S.~Walker,
  ``Ion-dependent dna configuration in bacteriophage capsids,'' {\em
  Biophysical Journal}, vol.~120, no.~16, pp.~3292--3302, 2021.

\bibitem{Qiu12}
X.~Qiu, ``Heat induced capsid disassembly and dna release of bacteriophage
  $\lambda$,'' {\em PLOS ONE}, vol.~7, pp.~1--6, 07 2012.

\bibitem{Cer22}
P.~Cermelli and G.~Indelicato, ``Coarse-grained mechanical models for viral
  capsids,'' {\em International Journal of Non-Linear Mechanics}, vol.~145,
  p.~104112, 2022.

\bibitem{gennes}
P.~G. de~Gennes, {\em Scaling Concepts in Polymer Physics}.
\newblock Ithaca: Cornell University Press, 1979.

\bibitem{pol_phys}
M.~Rubinstein and C.~Ralph~H., {\em Polymer Physics}, vol.~1.
\newblock New York: Oxford University Press, 2003.

\bibitem{int_pol_phys}
M.~Doi, {\em Introduction to Polymer Physics}.
\newblock New York: Oxford University Press, 1996.

\bibitem{Lev04}
Y.~Levin, M.~A. Idiart, and J.~J. Arenzon, ``Solute diffusion out of a
  vesicle,'' {\em Physica A: Statistical Mechanics and its Applications},
  vol.~344, no.~3, pp.~543 -- 546, 2004.

\bibitem{Idi04}
M.~A. Idiart and Y.~Levin, ``Rupture of a liposomal vesicle,'' {\em Phys. Rev.
  E}, vol.~69, p.~061922, 2004.

\bibitem{Levin04}
Y.~Levin and M.~A. Idiart, ``Pore dynamics of osmotically stressed vesicles,''
  {\em Physica A: Statistical Mechanics and its Applications}, vol.~331, no.~3,
  pp.~571 -- 578, 2004.

\bibitem{And50}
T.~F. Anderson, ``Destruction of bacterial viruses by osmotic shock,'' {\em J.
  App. Phys.}, vol.~21, p.~70, 1950.

\bibitem{Lei66}
L.~S. P. and M.~P., ``Effect of osmotic shock and low salt concentration on
  survival and density of bacteriophages t4b and t4bo1,'' {\em Biophys J.},
  vol.~6, p.~747, 1966.

\bibitem{Fred05}
G.~Fredrickson, {\em {The Equilibrium Theory of Inhomogeneous Polymers}}.
\newblock Oxford University Press, 12 2005.

\bibitem{Zha04}
D.~Zhang, R.~Konecny, N.~A. Baker, and J.~A. McCammon, ``Electrostatic
  interaction between rna and protein capsid in cowpea chlorotic mottle virus
  simulated by a coarse-grain rna model and a monte carlo approach,'' {\em
  Biopolymers}, vol.~75, no.~4, pp.~325--337, 2004.

\bibitem{Men11}
D.~Meng, R.~Hjelm, J.~Hu, and J.~Wu, ``A theoretical model for the dynamic
  structure of hepatitis b nucleocapsid,'' {\em Biophysical Journal}, vol.~101,
  no.~10, pp.~2476--2484, 2011.

\bibitem{Kim15}
J.~Kim and J.~Wu, ``A thermodynamic model for genome packaging in hepatitis b
  virus,'' {\em Biophysical Journal}, vol.~109, no.~8, pp.~1689--1697, 2015.

\bibitem{Zhe12}
Z.~Jin and J.~Wu, ``Density functional theory for encapsidated
  polyelectrolytes: A comparison with monte carlo simulation,'' {\em The
  Journal of Chemical Physics}, vol.~137, no.~4, p.~044905, 2012.

\bibitem{Chi13}
Q.~Chi, G.~Wang, and J.~Jiang, ``The persistence length and length per base of
  single-stranded dna obtained from fluorescence correlation spectroscopy
  measurements using mean field theory,'' {\em Physica A: Statistical Mechanics
  and its Applications}, vol.~392, no.~5, pp.~1072--1079, 2013.

\bibitem{Cam15}
J.~A. Campillo-Balderas, A.~Lazcano, and A.~Becerra, ``Viral genome size
  distribution does not correlate with the antiquity of the host lineages,''
  {\em Frontiers in Ecology and Evolution}, vol.~3, 2015.

\bibitem{Chai19}
K.~V. Chaitanya, {\em Structure and Organization of Virus Genomes}, pp.~1--30.
\newblock Singapore: Springer Singapore, 2019.

\bibitem{Los12}
A.~Lo\v{s}dorfer~Bo\v{z}i\v{c}, A.~\v{S}iber, and R.~Podgornik, ``How simple
  can a model of an empty viral capsid be? charge distributions in viral
  capsids,'' {\em Journal of Biological Physics}, vol.~38, pp.~657--671, 2012.

\bibitem{Tam98}
M.~Tamashiro, Y.~Levin, and M.~Barbosa, ``Donnan equilibrium and the osmotic
  pressure of charged colloidal lattices,'' {\em Physics of Condensed Matter},
  vol.~1, pp.~337--343, 02 1998.

\bibitem{Odi03}
T.~Odijk and F.~Slok, ``Nonuniform donnan equilibrium within bacteriophages
  packed with dna,'' {\em The Journal of Physical Chemistry B}, vol.~107,
  no.~32, pp.~8074--8077, 2003.

\bibitem{Tin11}
C.~L. Ting, J.~Wu, and Z.-G. Wang, ``Thermodynamic basis for the genome to
  capsid charge relationship in viral encapsidation,'' {\em Proceedings of the
  National Academy of Sciences}, vol.~108, no.~41, pp.~16986--16991, 2011.

\bibitem{Scho13}
P.~van~der Schoot and R.~Zandi, ``Impact of the topology of viral rnas on their
  encapsulation by virus coat proteins,'' {\em Journal of Biological Physics},
  vol.~39, no.~2, pp.~289--299, 2013.

\bibitem{Schwab09}
D.~Schwab and R.~F. Bruinsma, ``Flory theory of the folding of designed rna
  molecules,'' {\em The Journal of Physical Chemistry B}, vol.~113, no.~12,
  pp.~3880--3893, 2009.

\bibitem{Roya09}
R.~Zandi and P.~{van der Schoot}, ``Size regulation of ss-rna viruses,'' {\em
  Biophysical Journal}, vol.~96, no.~1, pp.~9--20, 2009.

\bibitem{Los13}
A.~Lo\v{s}dorfer~Bo\v{z}i\v{c}, A.~\v{S}iber, and R.~Podgornik, ``Statistical
  analysis of sizes and shapes of virus capsids and their resulting elastic
  properties,'' {\em Journal of Biological Physics}, vol.~639, pp.~215--228,
  2013.

\bibitem{Blaa04}
R.~Blaak, S.~Auer, D.~Frenkel, and H.~L\"owen, ``Crystal nucleation of
  colloidal suspensions under shear,'' {\em Phys. Rev. Lett.}, vol.~93,
  p.~068303, 2004.

\bibitem{Anv18}
A.~L. Bo\v{z}i\v{c} and A.~\v{S}iber, ``Electrostatics-driven inflation of
  elastic icosahedral shells as a model for swelling of viruses,'' {\em
  Biophysical Journal}, vol.~115, pp.~822 -- 829, 2018.

\end{thebibliography}

\end{document}